# A Metadynamics-Based Framework for Free Energy Surface Mapping of Multiparticle Diffusion in Crystals

Shunya Yamada* and Kazuaki Toyoura**

*Department of Materials Science and Engineering, Kyoto University, Kyoto 606-8501, Japan*

*yamada.shunya.73m@st.kyoto-u.ac.jp

**toyoura.kazuaki.5r@kyoto-u.ac.jp

We propose two metadynamics (MetaD)-based methodologies for efficiently mapping free energy surfaces (FESs) of multiple interacting carriers diffusing in crystalline solids. Our approaches circumvent the challenges of high-dimensional collective variables (CVs) by employing replica state exchange MetaD and parallel bias MetaD, both of which decompose the high-dimensional CVs into multiple lower-dimensional ones. As a benchmark, we investigate two-dimensional lithium (Li) diffusion in $Li_xTiS_2$ across a wide range of Li concentrations. The Li jump frequencies estimated from the obtained FESs exhibit concentration- and temperature-dependent trends consistent with previous kinetic Monte Carlo simulations and nuclear magnetic resonance measurements. Our approaches provide a promising framework for capturing the slow dynamics of diffusive carriers that are typically inaccessible to conventional molecular dynamics simulations.

Atomic diffusion in solids plays a critical role in various phenomena related to physics, chemistry, and materials science [1,2]. Theoretical analyses using first-principles calculations are powerful tools for understanding the dynamics of diffusion carriers at the atomic scale, for which "the Nudged Elastic Band (NEB) method [3,4] with the kinetic Monte Carlo (KMC) method [5–7]" and "the Molecular Dynamics (MD) method [8–10]" have been widely used so far.

Although the two conventional methods have achieved significant success for several decades, both methods possess individual inherent limitations. The former requires prior knowledge on the initial and final sites of every atomic jump in a system of interest, making it difficult to apply to systems with unknown diffusion mechanisms. Particularly, in the case of multiparticle diffusion, the potential energy barriers for atomic jumps depend on the local configuration of other diffusion carriers, rendering the analysis more challenging. Although the cluster expansion-based modeling of potential energy barriers [11,12] has been proposed as an elegant method to evaluate the local configuration dependence, it is not widely used probably due to the technical difficulties in the model construction.

In contrast, the latter method does not encounter such difficulties even in multiparticle diffusion because the complete trajectories of all diffusion carriers are obtained. However, MD simulations have a critical issue related to the accessible time scales. Specifically, the time step must be sufficiently shorter than the time scale of lattice vibration, while the total simulation time must be long comparable to the diffusion time scale, leading to a huge number of simulation steps. This time-scale requirement significantly limits the applicability of MD simulations.

The metadynamics (MetaD) method [13,14] is an effective technique for addressing the time-scale limitations of MD simulations, which has been developed primarily for analyzing structure dynamics and chemical reactions of molecular systems such as protein folding. In this method, Gaussian-shaped small bias potentials (Gaussian hills) are sequentially deposited in a predefined collective variable (CV) space, thereby accelerating the time evolution of the system. Since the accumulated Gaussian hills reflect the shape of the underlying free energy surface (FES) in the CV space, the FES can be reconstructed from the final bias potential after the simulation. The *multi-hill* MetaD method for interstitial diffusion has recently been proposed [15], in which multiple Gaussian hills are deposited simultaneously by exploiting the crystallographic symmetry (See Figs. 1a and 1b for the comparison between the conventional single-hill MetaD and the multi-hill MetaD). In this method, the relative coordinate of an interstitial atom with respect to the host crystal is chosen as the CV space. The reconstructed FES contains the detailed information on all stable and metastable interstitial sites, migration paths, and their free energy barriers. Based on the transition state theory (TST) [16], the atomic jump frequencies along these migration paths can be quantitatively evaluated, which enable estimating the diffusion coefficient by the KMC simulations. Notably, this method does not require any prior knowledge on the diffusion mechanism, and the estimated jump frequencies takes anharmonic lattice dynamics into account, which are clear advantages over the NEB method.

Nevertheless, the applicability of the multi-hill MetaD method is limited to single-particle diffusion by the interstitial mechanism, where the CV space has three dimensions at most. When straightforwardly expanded to multiparticle diffusion, the CV space becomes higher dimensional, i.e., a $3N$-dimensional CV space in an $N$-carrier system. In practice, the multi-hill MetaD is no longer effective in the CV space beyond three dimensions, because the computational cost exponentially increases with the dimensionality of the CV space well-known as *the curse of dimensionality* [17].

In this letter, we propose a feasible MetaD-based framework of the FES mapping for multiparticle diffusion. Specifically, we employ the Replica State Exchange MetaD (RSE-MetaD) method [18,19] and the Parallel Bias MetaD (PB-MetaD) method [20–22], both of which have the potential to overcome the curse of dimensionality by decomposing a high-dimensional CV space into multiple lower-dimensional CV spaces. We demonstrate the efficacy of RSE-MetaD and PB-MetaD combined with the multi-hill MetaD method for multiparticle diffusion, using lithium-intercalated titanium disulfide $Li_xTiS_2$ $(0 \leq x \leq 1)$ as a model system, where Li atoms undergo two-dimensional diffusion in the $TiS_2$ interlayers.

Before describing the details of the proposed methodology, the concept of the two approaches based on RSE-MetaD and PB-MetaD are outlined hereafter. Figures 1c and 1d show the schematic illustrations of the Gaussian hill deposition in RSE-MetaD and PB-MetaD for a $N$-carrier system, respectively. In RSE-MetaD, multi-hill MetaD simulations using $N$ replicas are performed in parallel, in each of which the coordinate of a different carrier is used as the CVs. These simulations are not independent; the replica states are exchanged at predefined time intervals based on the replica exchange scheme to enhance the sampling efficiency. When the diffusion carriers are of the same species, the *single-particle* FES can be reconstructed by averaging the bias potentials in all CV spaces. Even in the case of multiple carrier species, the single-particle FES of each species can be reconstructed readily by averaging within the same species. The interactions with other diffusion carriers are incorporated into the obtained single-particle FES as a potential of mean force. In PB-MetaD, only a single multi-hill MetaD simulation is performed, in which Gaussian hills are simultaneously deposited in all CV spaces. This strategy offers a significant computational advantage compared to RSE-MetaD requiring $N$ parallel simulations. Note that the height of Gaussian hills in each CV space is weighted from the requirements of statistical mechanics, as detailed below. The single-particle FES can be reconstructed by averaging the bias potentials in all CV spaces within the same species, as in RSE-MetaD.

The proposed methodologies are detailed below. In the conventional MetaD for single-particle diffusion, the bias potential $V^{\text{bias}}(n; \boldsymbol{X})$ at the $n$-th deposition step is given by

$$V^{\text{bias}}(n; \boldsymbol{X}) = V^{\text{bias}}(n-1; \boldsymbol{X}) + h_n \exp\left[-\frac{\|\boldsymbol{X} - \boldsymbol{X}_n\|^2}{2w^2}\right], \tag{1}$$

where $\boldsymbol{X}$ denotes the relative coordinate of the single carrier with respect to the host crystal, $\boldsymbol{X}_n$ is $\boldsymbol{X}$ at the $n$-th deposition step, and $h_n$ and $w$ are the height and width of the Gaussian hill, respectively. In the well-tempered MetaD scheme [23], $h_n$ is given by

$$h_n = h_0 \exp\left[-\frac{V^{\text{bias}}(n-1;\boldsymbol{X}_n)}{k_\text{B}\Delta T}\right], \tag{2}$$

where $h_0$ is the initial Gaussian height, $k_\text{B}$ is the Boltzmann constant, and $\Delta T$ is the temperature-valued parameter controlling the decay of the Gaussian height. In multi-hill MetaD, Eq. (1) is converted into

$$V^{\text{bias}}(n;\boldsymbol{X}) = V^{\text{bias}}(n-1;\boldsymbol{X}) + \sum_m h_n \exp\left[-\frac{\|\boldsymbol{X} - \mathbf{T}_m\boldsymbol{X}_n\|^2}{2w^2}\right], \tag{3}$$

where $\mathbf{T}_m$ represents the *m*-th symmetry operation generating crystallographically equivalent points of $\boldsymbol{X}_n$. Under the converged condition, the free energy $F(\boldsymbol{X})$ is reconstructed by

$$F(\boldsymbol{X}) = -(1 + T/\Delta T)V^{\text{bias}}(\boldsymbol{X}). \tag{4}$$

In RSE-MetaD for *N*-particle diffusion, *N* multi-hill MetaD simulations are performed in parallel. In the *i*-th simulation, the coordinate of *i*-th carrier, $\boldsymbol{X}^{(i)}$ $(i = 1,2,\dots,N)$, is chosen as the CVs, where the bias potential, $V_i^{\text{RSE}}(\boldsymbol{X}^{(i)})$, acts only on the *i*-th carrier and has no effect on the other carriers. The coordinates and velocities of all atoms (*replica states*) are exchanged between these simulations at predefined time intervals, for which several exchange schemes have been proposed so far [18,24,25]. The proposed approach employs the *infinite swapping* (INS) method [25] considering all permutations of *N* replica states, i.e., *N*! possibilities. Considering the replica state exchange from an initial state $\boldsymbol{\sigma} = (\sigma_1, \sigma_2, \dots, \sigma_N)$ to a permutation $\boldsymbol{\sigma}' = (\sigma'_1, \sigma'_2, \dots, \sigma'_N)$, the permutation $\boldsymbol{\sigma}'$ is selected from the *N*! possible permutations according to the following statistical weight,

$$W_{\boldsymbol{\sigma}'}^{\text{RSE}} \propto \prod_{i=1}^{N} \exp\left[-\frac{V_i^{\text{RSE}}\left(\boldsymbol{X}_{\sigma'_i}^{(i)}\right)}{k_\text{B}T}\right], \tag{5}$$

where $\boldsymbol{X}_{\sigma'_i}^{(i)}$ denotes the coordinate of *i*-th carrier in the *i*-th replica state in the permutation $\boldsymbol{\sigma}'$. The INS method has the advantages of higher exchange probability and higher sampling efficiency than the conventional exchange scheme between randomly-chosen replica-state pairs based on the Metropolis criterion as in the bias exchange metadynamics (BE-MetaD) [24]. In the case of large *N*, the partial INS (pINS) method [25] is employed to mitigate the huge number of permutations. Under the converged conditions, the single-particle FES, $F^{\mathrm{SP}}(\boldsymbol{X})$, can be reconstructed as follows:

$$F^{\mathrm{SP}}(\boldsymbol{X}) = -\left(1 + \frac{T}{\Delta T}\right)\frac{1}{N}\sum_{i=1}^{N} V_i^{\mathrm{RSE}}(\boldsymbol{X}). \tag{6}$$

In PB-MetaD, the bias potential $V_i^{\mathrm{PB}}(n;\boldsymbol{X})$ acting on the $i$-th carrier at the $n$-th deposition step is given by

$$V_i^{\mathrm{PB}}(n;\boldsymbol{X}) = V_i^{\mathrm{PB}}(n-1;\boldsymbol{X}) + \sum_m \alpha_{i,n} h_{i,n} \exp\left[-\frac{\left\|\boldsymbol{X} - \mathrm{T}_m \boldsymbol{X}_n^{(i)}\right\|^2}{2w^2}\right], \tag{7}$$

where $\boldsymbol{X}_n^{(i)}$ is the coordinate of the $i$-th carrier at the $n$-th deposition step, and $\alpha_{i,n}$ is the weight factor defined by

$$\alpha_{i,n} = \frac{\exp\left[-V_i^{\mathrm{PB}}\left(n-1;\boldsymbol{X}_n^{(i)}\right)/k_{\mathrm{B}}T\right]}{\sum_{i=1}^{N}\exp\left[-V_i^{\mathrm{PB}}\left(n-1;\boldsymbol{X}_n^{(i)}\right)/k_{\mathrm{B}}T\right]}. \tag{8}$$

The individual Gaussian height $h_{i,n}$ is given by the following formula analogous to Eq. (2):

$$h_{i,n} = h_0 \exp\left[-\frac{V_i^{\mathrm{PB}}\left(n-1;\boldsymbol{X}_n^{(i)}\right)}{k_{\mathrm{B}}\Delta T}\right]. \tag{9}$$

Note that the unified bias potential $V_{\mathrm{uni}}^{\mathrm{PB}}$ to all carriers is not the simple sum of $V_i^{\mathrm{PB}}$, but is given by

$$V_{\mathrm{uni}}^{\mathrm{PB}}\left(\boldsymbol{X}^{(1)},\boldsymbol{X}^{(2)},\ldots,\boldsymbol{X}^{(N)}\right) = -k_{\mathrm{B}}T\ln\left[\sum_{i=1}^{N}\exp\left[-V_i^{\mathrm{PB}}\left(\boldsymbol{X}^{(i)}\right)/k_{\mathrm{B}}T\right]\right]. \tag{10}$$

From Eq. (10), the bias force acting on the $i$-th carrier $f_i^{\mathrm{PB}}$ is given by

$$f_i^{\mathrm{PB}} = -\frac{\partial V_{\mathrm{uni}}^{\mathrm{PB}}\left(\boldsymbol{X}^{(1)},\boldsymbol{X}^{(2)},\ldots,\boldsymbol{X}^{(N)}\right)}{\partial \boldsymbol{X}^{(i)}} = -\frac{\exp\left[-V_i^{\mathrm{PB}}\left(\boldsymbol{X}^{(i)}\right)/k_{\mathrm{B}}T\right]}{\sum_{i=1}^{N}\exp\left[-V_i^{\mathrm{PB}}\left(\boldsymbol{X}^{(i)}\right)/k_{\mathrm{B}}T\right]}\frac{\partial V_i^{\mathrm{PB}}\left(\boldsymbol{X}^{(i)}\right)}{\partial \boldsymbol{X}^{(i)}}. \tag{11}$$

This relation implies that the force derived from the bias is also weighted according to the relative magnitude of each carrier's bias. The single-particle FES can be reconstructed using the following formula analogous to Eq. (6);

$$F^{\mathrm{SP}}(\boldsymbol{X}) = -\left(1+\frac{T}{\Delta T}\right)\frac{1}{N}\sum_{i=1}^{N} V_i^{\mathrm{PB}}(\boldsymbol{X}). \tag{12}$$

Once the single-particle FES is obtained, all elementary processes can be listed readily by a path search algorithm on the FES. The atomic jump frequency of elementary process $k$, $\Gamma_k$, can be evaluated based on the TST as

$$\Gamma_k = \sqrt{\frac{k_{\mathrm{B}}T}{2\pi M}}\frac{\iint_{\mathrm{S}_k} \exp\left[-\frac{F^{\mathrm{SP}}(\boldsymbol{X})}{k_{\mathrm{B}}T}\right] dS}{\iiint_{\mathrm{V}_k} \exp\left[-\frac{F^{\mathrm{SP}}(\boldsymbol{X})}{k_{\mathrm{B}}T}\right] dV}, \tag{13}$$

where $M$ is the carrier mass. The numerator and denominator correspond to the surface integral over the transition-state surface, $\mathrm{S}_k$, and the volume integral over the vicinity of the initial state, $\mathrm{V}_k$, respectively. The diffusion coefficient can be estimated by either the KMC simulations or numerical solution of the master equation [26]. It should be noted, however, that the dynamical correlations, i.e., both auto-correlation and cross-correlation, are neglected in principle. This is because the obtained diffusion coefficient corresponds to that of an apparent single particle moving on the FES incorporating the interactions with the other carriers as a potential of mean force. Although these dynamical correlation effects should ideally be estimated from the trajectories of diffusion atoms in MD simulations with no applied bias potential, it is also possible to approximate them using the atomic trajectories in MetaD simulations. In this paper, we discuss the methodology and validity of this post-processing technique.

In the present study, the proposed approaches based on RSE-MetaD and PB-MetaD were carried out using the on-the-fly machine learning force field (MLFF) [27,28] implemented in the Vienna Ab initio Simulation Package (VASP) code [29–31]. A supercell composed of 4×4×1 unit cells of $TiS_2$ (space group: $P\bar{3}m1$) with $N$ Li atoms intercalated into the interlayers ($Li_NTi_{16}S_{32}$, $1 \leq N \leq 15$) was employed for the MetaD simulations. The CVs were defined as the two-dimensional coordinates of each Li atom in the $TiS_2$ interlayer. Gaussian hills were deposited every 50 MD steps at 192 equivalent positions according to the multi-hill strategy. The initial height, width, and decay factor of the Gaussian hill ($h_0$, $w$, $k_B\Delta T$) were set to 0.02 eV, 0.1 Å, and 0.3 eV, respectively. The MD simulations were performed in the *NPT* ensemble with the fixed *ab*-axes and the variable *c*-axis length. The other computational details and the accuracies of the constructed MLFFs are provided in Secs. S1 and S2, respectively, of the supplementary materials [32]. The Li diffusion coefficients were evaluated via numerical solution of the master equation using the MASTEQ code [26].

Figure 2a shows the crystal structure of $Li_xTiS_2$, in which S atoms form a hexagonal close-packed sublattice, and Ti atoms occupy the octahedral sites in every other layer along the *c*-axis. Li atoms are intercalated between the $TiS_2$ layers and undergo two-dimensional diffusion in the interlayers shown in Fig. 2b. According to the previous study based on Nuclear Magnetic Resonance (NMR) measurements [33] and first-principles calculations [12], Li atoms occupy the octahedral sites and migrate between adjacent octahedral sites via an intermediate tetrahedral site.

Figure 2c shows the two-dimensional Li-FESs at $x$ = 2/16, 8/16, and 15/16 in $Li_xTiS_2$ evaluated by RSE-MetaD and PB-MetaD at 300 K (See Sec. S3 in the supplementary materials [32] for the Li-FESs evaluated at all compositions). Except for $x$ = 15/16, the global and local minima of the Li-FESs are located at the octahedral sites and the tetrahedral sites, respectively, which is consistent with the previous theoretical and experimental studies [12,33]. Although the tetrahedral site is not a local minimum at $x$ = 15/16, this observation is consistent with the fact that tetrahedral sites adjacent to a single vacancy form shallower local minima (15 meV) than those adjacent to two neighboring vacancies, i.e. a *divacancy* (150 meV) [12]. At $x$ = 2/16, 8/16, and 15/16, the evaluated free energy barriers from an octahedral site to an adjacent tetrahedral site are 0.67, 0.49 and 0.70 eV via RSE-MetaD and 0.68, 0.48, 0.71 eV via PB-MetaD, respectively. This indicates that the similar Li-FESs can be obtained by the two proposed approaches. The free energy barrier is minimal at intermediate Li compositions, which can be qualitatively understood from the change in $TiS_2$ interlayer distance and the site blocking effect by other Li atoms. As previously reported [12], the $TiS_2$ interlayer distance monotonically increases with Li composition $x$ and reaches a plateau above the intermediate region, which is also observed in the present study (see Sec. S4 in the supplementary materials [32]). This interlayer extension significantly decreases the Li migration barriers, resulting in a decrease in the free energy barrier with $x$ in the low-$x$ region. At high Li compositions, the site blocking effect by other Li atoms occupying adjacent sites increases with Li composition, leading to the increased free energy barrier. In addition, the relatively high potential energy barrier of Li jumps into single vacancies should also raise the free energy barrier at high Li compositions. This implies that the obtained FESs by the two proposed approaches should reflect the local environment of Li atoms varying with Li configurations.

At the Li composition $x$ = 11/16 (~ 0.7), the RSE-MetaD and PB-MetaD simulations were performed over a wide temperature range from 150 to 1500 K for comparison with the Li jump frequencies at $x$ = 0.7 measured by the spin lattice relaxation and spin alignment echo NMR in the literature [33]. Figure 3 shows the Arrhenius plots of the Li jump frequencies, in which the red solid squares, black solid circles, blue solid diamonds, and green open triangles correspond to the RSE-MetaD, the PB-MetaD, the conventional MD, and the NMR, respectively. These apparent activation energies are 0.47, 0.48, 0.50, and 0.41 eV, respectively. The estimated jump frequencies by the three computational approaches are in excellent agreement with each other, but have slightly higher activation energies than those of the NMR measurements. This discrepancy is mainly attributed to the relatively short *ab*-axes in our simulations to the experimental values determined by the X-ray diffraction at room temperature [34]. In the present study, the fixed *ab*-axes were used in all MD and MetaD simulations, which were optimized in advance by first-principles calculations under the local density approximation (LDA). The *ab*-axes lengths are approximately 3 % shorter than the experimental lattice constants (3.43 Å for $x$ = 0.70 at room temperature) primarily due to the thermal expansion effects and the systematic error inherent in the LDA. The additional PB-MetaD simulations were, therefore, performed with the *ab*-axes lengths fixed at the experimental values (black open circles in Fig. 3), whose apparent activation energy (0.43 eV) became significantly closer to the NMR value. Generally, the TST (Eq. (13)) tends to overestimate jump frequencies at high temperatures or when the transition state cannot be defined clearly, owing to dynamical recrossing [35]. For this model system, however, the transmission coefficient estimated from the conventional MD simulation was close to unity even at 1500 K (See Sec. S5 in the supplementary materials [32]), indicating that the dynamical recrossing effect is negligible.

Figure 4 shows the Li diffusion coefficients at 300 K as a function of Li composition, in which the red squares and the black circles denote the Li diffusion coefficients evaluated by the RSE-MetaD and the PB-MetaD simulations ($D_{\mathrm{RSE}}$ and $D_{\mathrm{PB}}$), respectively. For $Li_{1/16}TiS_2$ (shown by the gray triangles), the normal multi-hill MetaD simulation was performed because there is only a single Li atom in the simulation cell. Reflecting the low free energy barriers at intermediate Li compositions, $D_{\mathrm{RSE}}$ and $D_{\mathrm{PB}}$ are unimodal functions of Li composition with a maximum at an intermediate composition, which is in qualitative agreement with the tracer diffusion coefficients $D^{*}_{\mathrm{KMC}}$ (green diamonds) estimated using the KMC simulations with the local cluster expansion technique [12]. However, $D_{\mathrm{RSE}}$ and $D_{\mathrm{PB}}$ in the present study are higher than $D^{*}_{\mathrm{KMC}}$ by two orders of magnitude in intermediate Li compositions from the quantitative point of view. In Ref. [12], the authors mentioned that their predictions tend to underestimate experimental values [36] by a few orders of magnitude, meaning that our predictions more accurately reproduce the experimental diffusivity.

A possible origin of the discrepancy of $D_{\mathrm{RSE}}$ and $D_{\mathrm{PB}}$ from $D^{*}_{\mathrm{KMC}}$ is the difference in the $c$-axis length of the simulation cell. It seems that the previous KMC study neglected $c$-axis thermal expansion at finite temperatures. Our additional validation (See Sec. S6 in the supplementary materials [32]) indicates that the thermal expansion of $c$-axis length 5–10-fold enhances Li diffusivity. Another possible origin is the loss of information on the dynamical correlation in the single-particle FES obtained via RSE-MetaD and PB-MetaD. $D^{*}_{\mathrm{KMC}}$ includes the auto-correlation effect, while $D_{\mathrm{RSE}}$ and $D_{\mathrm{PB}}$ do not include such a correlation effect at all, leading to the overestimation of Li diffusivity in our proposed framework. However, the overestimation is only about two-fold based on the Li trajectories in our PB-MetaD simulations, suggesting that the former origin should be the primary factor of the discrepancy. See Sec. S7 in the supplementary materials [32] for more quantitative discussion on the effect of the dynamical correlation factor on the lithium diffusivity, in which the post-processing technique for estimating the dynamical correlation factors from the atomic trajectories in MetaD simulations and its validity are discussed.

The error bars in Fig. 4 indicate the convergence error of the FES with respect to the simulation time. The bar length corresponds to the double standard deviation $2\sigma_{\mathrm{std}}$ of $\log D$ sampled at every 1 ps in the last 100 ps. The $\log D$ profiles at all Li compositions are provided in Sec. S8 in the supplementary materials [32]. For $x \leq 9/16$ in both RSE-MetaD and PB-MetaD, the convergence errors are 0.25 at most on the common logarithmic scale, corresponding to $k_{\mathrm{B}}T\ln(10^{0.25}) \sim 15$ meV at 300 K in the energy scale. In contrast, for $10/16 \leq x \leq 14/16$, the error becomes relatively large, e.g., $k_{\mathrm{B}}T\ln(10^{0.69}) \sim 41$ meV at $x = 14/16$ in PB-MetaD. This large error originates from the strong dependence of potential energy barrier on the local Li configuration around a migrating Li atom. Specifically, *divacancy* formation significantly reduces the potential energy barrier [12], which causes abrupt increase in the Li jump frequency. This abrupt change in the Li jump frequency substantially affects the free energy estimation, particularly in PB-MetaD which relies only on a single simulation. The most straightforward way to reduce the convergence errors in PB-MetaD is averaging the FESs in multiple independent simulations. Actually, $D_{\mathrm{PB}}$ estimated from the averaged FES in $N$ PB-MetaD simulations (blue circles in Fig. 4) have negligible convergence errors. Only at $x$ = 14/16, $D_{\mathrm{PB}}$ and $D_{\mathrm{RSE}}$ differ by one order of magnitude after this averaging, which exceeds the range of convergence errors. This discrepancy stems from the persistent heterogeneity of local environments experienced by individual biased Li atoms in RSE-MetaD, caused by slow replica state exchanges involving the *divacancy* configuration. A detailed discussion is provided in Sec. S9 of the supplementary materials [32].

Concerning the computational costs of the proposed methodology, both RSE-MetaD and PB-MetaD approaches provide substantial advantages over the conventional MD. Figure 5 shows the estimated number of MD steps until $10N$ Li jumps are observed in a conventional MD simulation, $n_{\mathrm{MD}}$, as a function of Li composition $x$ and temperature $T$, which is roughly estimated from the Li jump frequency and the free-energy barrier evaluated by PB-MetaD at 300 K. Reflecting the relatively high Li jump frequency at intermediate Li compositions, $n_{\mathrm{MD}}$ is a unimodal function of Li composition with a minimum at an intermediate composition under the constant temperature condition. Under the constant Li composition condition, $n_{\mathrm{MD}}$ is a monotonically decreasing function of temperature, because the Li jump is a thermally activated process. By contrast, the number of MD steps required by both RSE-MetaD and PB-MetaD, $n_{\mathrm{RSE}}$ and $n_{\mathrm{PB}}$, are nearly constant with respect to the temperature (see Sec. S10 in the supplementary materials [32]). RSE-MetaD requires $N$ parallel simulations for each carrier in the simulation cell, meaning that PB-MetaD is more efficient than RSE-MetaD ($n_{\mathrm{RSE}} = N \times n_{\mathrm{PB}}$). The red, green, and blue bars in Fig. 5 indicate $n_{\mathrm{RSE}} \leq n_{\mathrm{MD}}$, $n_{\mathrm{PB}} \leq n_{\mathrm{MD}} < n_{\mathrm{RSE}}$, and $n_{\mathrm{MD}} < n_{\mathrm{PB}}$, respectively, suggesting that RSE-MetaD and PB-MetaD is more efficient than the conventional MD in the low temperature range, particularly efficient at room temperature of practical importance.

In conclusion, we have proposed a MetaD-based framework for efficiently mapping the FES in multiparticle diffusion. The proposed approach offers a promising alternative for analyzing the slow dynamics of diffusive carriers in crystalline solids, which are often inaccessible to conventional MD simulations. Furthermore, the significant enhancement in computational efficiency achieved here is expected to facilitate large-scale materials exploration, such as the discovery of solid-state electrolytes, where extensive sampling is indispensable.

This work was supported by Precursory Research for Embryonic Science and Technology (PRESTO, Grant No. JPMJPR24J8) from Japan Science and Technology Agency (JST), and KAKENHI (Grants No. 24K01147 and No. 26KJ1521) from Japan Society for the Promotion of Science (JSPS).

## FIGURE CAPTIONS

FIG. 1. Schematic illustrations of MetaD-based methods for FES mapping of diffusion carriers, exemplified by a 2D supercell consisting of 2×2 unit cells. The square lattices in the bottom plane represent the host lattice, while the red, green and blue symbols denote the diffusion carriers. (a) Conventional MetaD and (b) multi-hill MetaD for single-particle diffusion, while (c) RSE-MetaD and (d) PB-MetaD for multiparticle diffusion. In conventional MetaD, a single Gaussian hill is deposited at the carrier position, while four Gaussian hills are simultaneously deposited by exploiting the translational symmetry in multi-hill MetaD. For multiparticle diffusion, this model system contains three diffusion carriers. In RSE-MetaD, three multi-hill MetaD simulations are executed in parallel, each utilizing the coordinate of a different carrier as the CVs. At predetermined intervals, the atomic configurations and velocities, i.e., replica states, are exchanged among these simulations. By contrast, a single multi-hill MetaD simulation is executed in PB-MetaD, where Gaussian bias potentials are deposited simultaneously across all CV spaces, with their heights scaled by the Boltzmann factor according to the bias potential at each carrier's position.

FIG. 2. (a) Crystal structure of $Li_xTiS_2$. The green, yellow, and red spheres represent Ti, S, and Li atoms, respectively. Li atoms are intercalated between the $TiS_2$ layers and undergo two-dimensional diffusion within the interlayers. The Li atoms are stabilized at the octahedral sites (shown as red octahedra) and migrate between these sites via an intermediate tetrahedral site outlined in blue. (b) Arrangement of octahedral and tetrahedral sites on the Li diffusion plane, viewed along the *c*-axis. Red circles and blue diamonds represent octahedral and tetrahedral sites, respectively. (c) FESs in the Li-diffusion plane obtained from RSE-MetaD and PB-MetaD simulations. The color map shows the free energy of Li, referenced to the most stable octahedral-site position.

FIG. 3. Arrhenius plot of the Li jump frequency in $Li_{11/16}TiS_2$. The red squares, black circles, blue diamonds, and green triangles represent the results of RSE-MetaD, PB-MetaD, standard MD simulations, and the NMR experiments [33], respectively. The black open symbols denote the results of the PB-MetaD simulations with the *ab*-axes lattice parameters constrained to the XRD experimental value (3.43 Å) [34], which is approximately 3% larger than the relaxed values.

FIG. 4. Li diffusion coefficient $D$ in $Li_xTiS_2$ as a function of Li concentration. The grey triangles, red squares, black circles, and green diamonds represent the results of multi-hill MetaD for single-particle diffusion, RSE-MetaD, PB-MetaD for multiparticle diffusion, and the tracer diffusion coefficient evaluated by KMC simulations [12], respectively. The blue diamonds were obtained from the FESs averaged over $N$ independent PB-MetaD simulations. The open green symbols represent the results for staged configurations, in which Li atoms segregate into alternating layers.

FIG. 5. The required number of MD time steps, $n_{\mathrm{MD}}$, for evaluating the Li jump frequency via standard MD simulations, as a function of Li concentration and temperature. Red, green, and blue bars indicate $n_{\mathrm{RSE}} \leq n_{\mathrm{MD}}$, $n_{\mathrm{PB}} \leq n_{\mathrm{MD}} < n_{\mathrm{RSE}}$, and $n_{\mathrm{MD}} < n_{\mathrm{PB}}$, respectively.

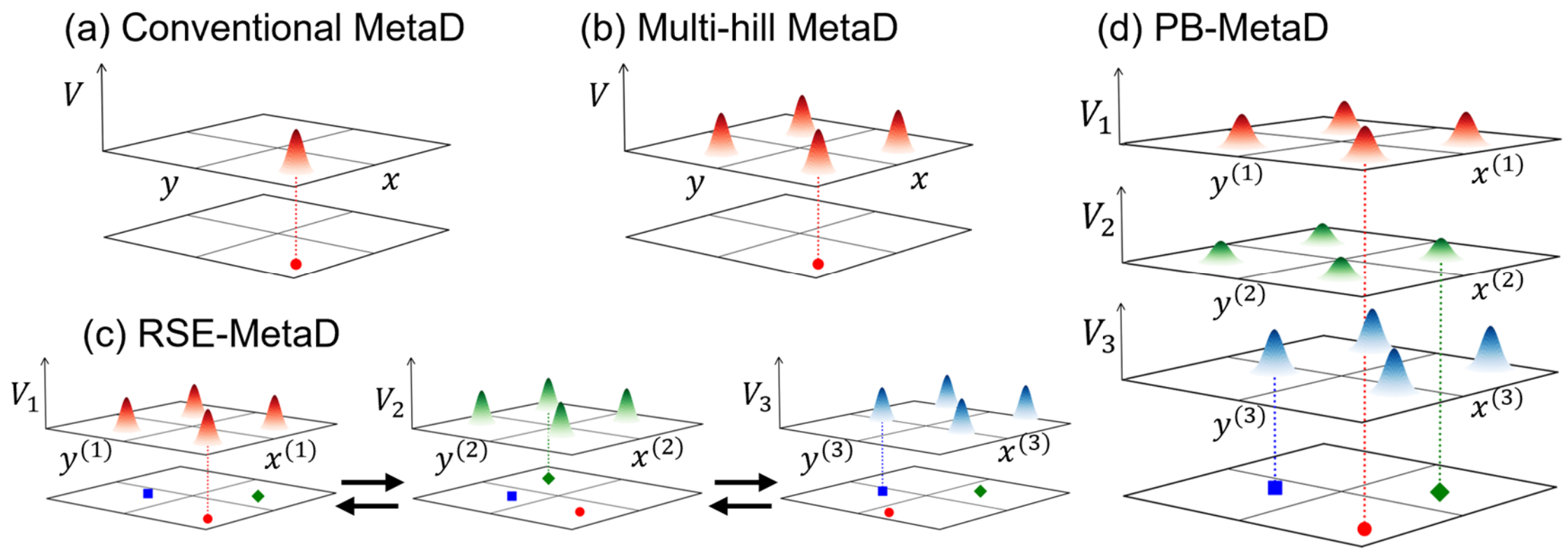

FIG. 1. Schematic illustrations of MetaD-based methods for FES mapping of diffusion carriers, exemplified by a 2D supercell consisting of 2×2 unit cells. The square lattices in the bottom plane represent the host lattice, while the red, green and blue symbols denote the diffusion carriers. (a) Conventional MetaD and (b) multi-hill MetaD for single-particle diffusion, while (c) RSE-MetaD and (d) PB-MetaD for multiparticle diffusion. In conventional MetaD, a single Gaussian hill is deposited at the carrier position, while four Gaussian hills are simultaneously deposited by exploiting the translational symmetry in multi-hill MetaD. For multiparticle diffusion, this model system contains three diffusion carriers. In RSE-MetaD, three multi-hill MetaD simulations are executed in parallel, each utilizing the coordinate of a different carrier as the CVs. At predetermined intervals, the atomic configurations and velocities, i.e., replica states, are exchanged among these simulations. By contrast, a single multi-hill MetaD simulation is executed in PB-MetaD, where Gaussian bias potentials are deposited simultaneously across all CV spaces, with their heights scaled by the Boltzmann factor according to the bias potential at each carrier's position.

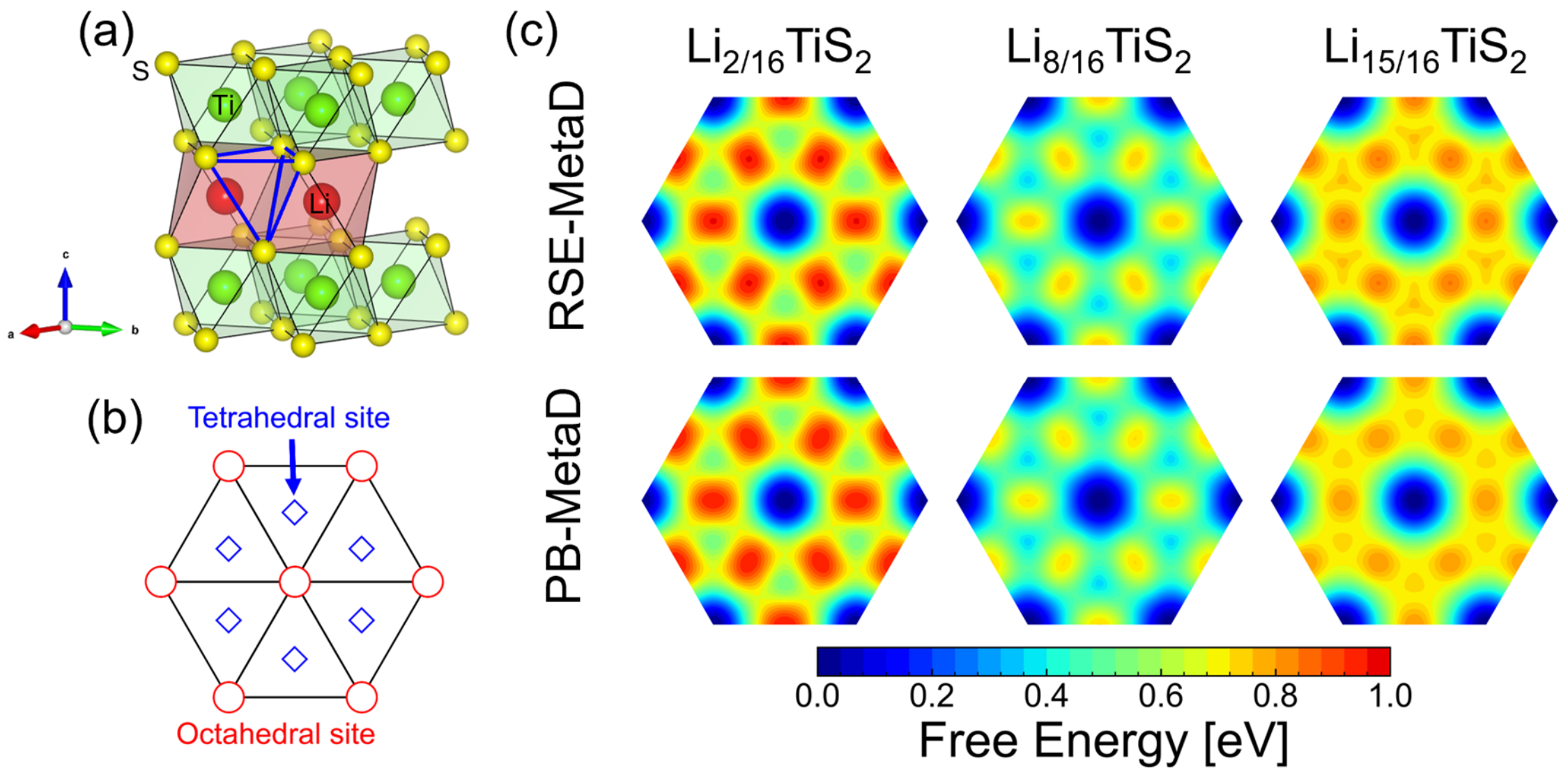

FIG. 2. (a) Crystal structure of $Li_xTiS_2$. The green, yellow, and red spheres represent Ti, S, and Li atoms, respectively. Li atoms are intercalated between the $TiS_2$ layers and undergo two-dimensional diffusion within the interlayers. The Li atoms are stabilized at the octahedral sites (shown as red octahedra) and migrate between these sites via an intermediate tetrahedral site outlined in blue. (b) Arrangement of octahedral and tetrahedral sites on the Li diffusion plane, viewed along the *c*-axis. Red circles and blue diamonds represent octahedral and tetrahedral sites, respectively. (c) FESs in the Li-diffusion plane obtained from RSE-MetaD and PB-MetaD simulations. The color map shows the free energy of Li, referenced to the most stable octahedral-site position.

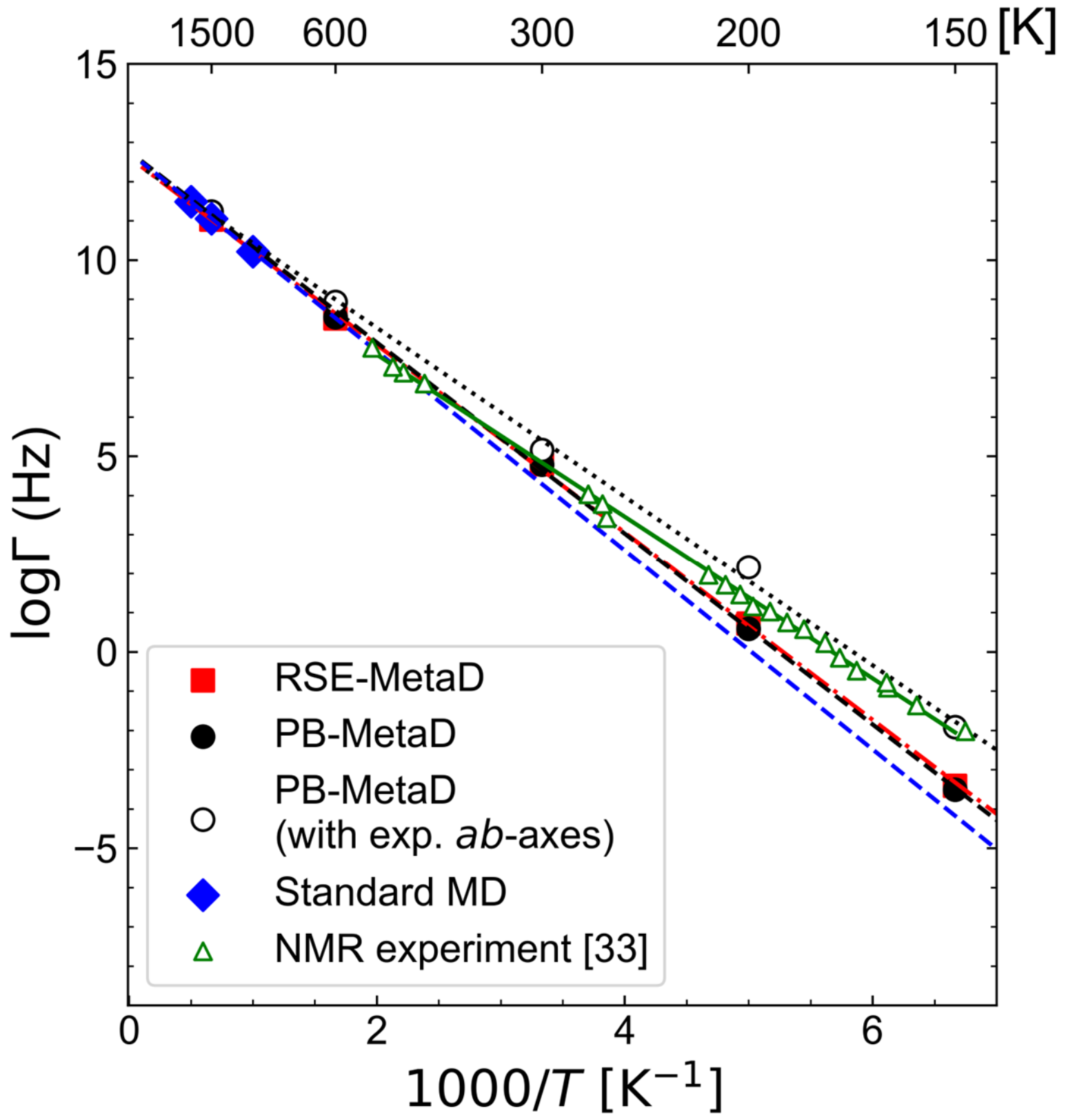

FIG. 3. Arrhenius plot of the Li jump frequency in $Li_{11/16}TiS_2$. The red squares, black circles, blue diamonds, and green triangles represent the results of RSE-MetaD, PB-MetaD, standard MD simulations, and the NMR experiments [33], respectively. The black open symbols denote the results of the PB-MetaD simulations with the *ab*-axes lattice parameters constrained to the XRD experimental value (3.43 Å) [34], which is approximately 3% larger than the relaxed values.

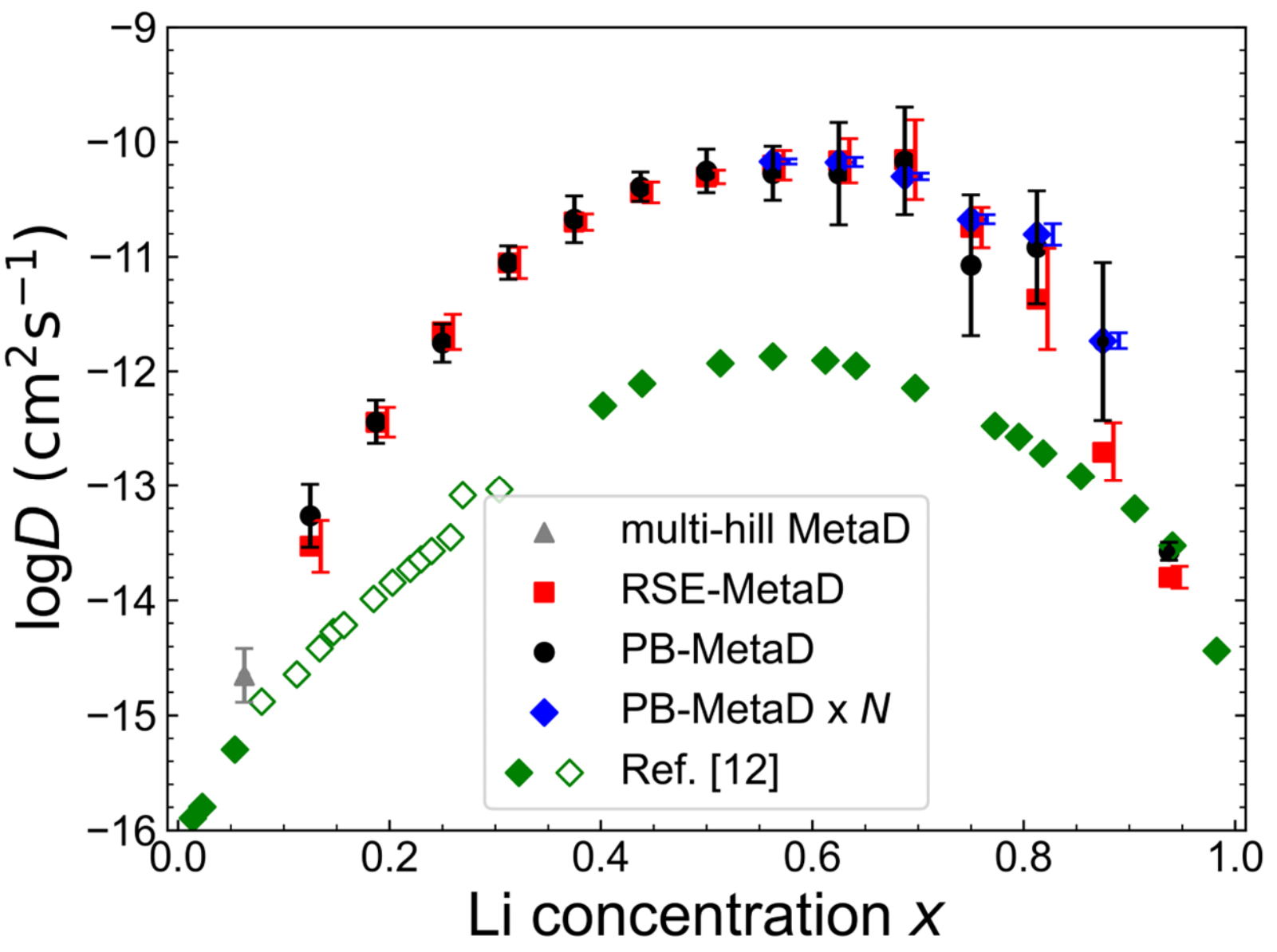

FIG. 4. Li diffusion coefficient $D$ in $Li_xTiS_2$ as a function of Li concentration. The grey triangles, red squares, black circles, and green diamonds represent the results of multi-hill MetaD for single-particle diffusion, RSE-MetaD, PB-MetaD for multiparticle diffusion, and the tracer diffusion coefficient evaluated by KMC simulations [12], respectively. The blue diamonds were obtained from the FESs averaged over $N$ independent PB-MetaD simulations. The open green symbols represent the results for staged configurations, in which Li atoms segregate into alternating layers.

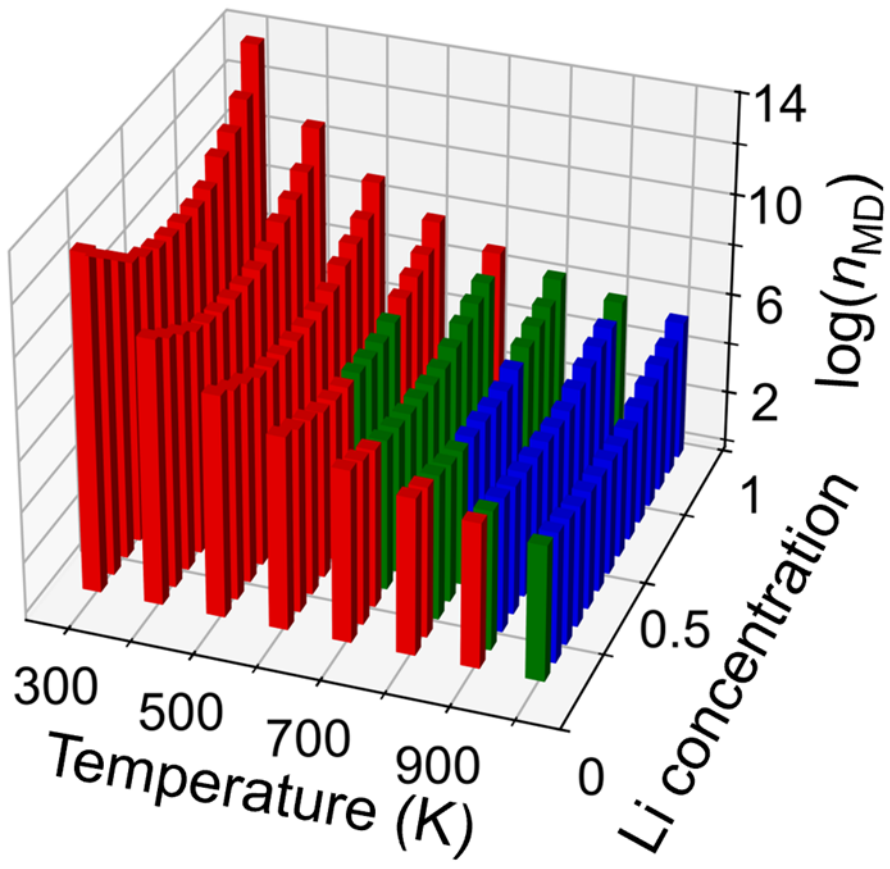

FIG. 5. The required number of MD time steps, $n_{\mathrm{MD}}$, for evaluating the Li jump frequency via standard MD simulations, as a function of Li concentration and temperature. Red, green, and blue bars indicate $n_{\mathrm{RSE}} \leq n_{\mathrm{MD}}$, $n_{\mathrm{PB}} \leq n_{\mathrm{MD}} < n_{\mathrm{RSE}}$, and $n_{\mathrm{MD}} < n_{\mathrm{PB}}$, respectively.

# *Supplementary Materials*

## A Metadynamics-Based Framework to Free Energy Surface Mapping for Multiparticle Diffusion in Crystals

Shunya Yamada* and Kazuaki Toyoura**

*Department of Materials Science and Engineering, Kyoto University, Kyoto 606-8501, Japan*

* yamada.shunya.73m@st.kyoto-u.ac.jp

** toyoura.kazuaki.5r@kyoto-u.ac.jp

### S1. Computational details

The MLFFs implemented in the VASP code employ the Smooth Overlap of Atomic Positions (SOAP) descriptor [36] combined with Gaussian-process regression to fit the total energy, atomic forces, and pressure. For simulations below 1000 K, additional DFT-based calculations were carried out to retrain the MLFFs whenever the regression error in forces for any atom exceeded 0.05 eV/Å to maintain the accuracy under accelerated sampling conditions. For simulations above 1000 K, we used the MLFFs constructed via on-the-fly training for $5\times10^4$ steps, with the temperature increasing linearly from 1500 K to 2500 K. DFT calculations were performed using the projector augmented wave (PAW) method [37,38] with the local density approximation (LDA) following the literature [12]. In the PAW potentials, the 3d and 4s orbitals for Ti, the 3s and 3p orbitals for S, and the 2s orbital for Li were treated as valence states. The plane-wave cutoff energy was set to 350 eV. We employed a supercell consisting of 4×4×1 unit cells of $TiS_2$ with $N$ Li atoms intercalated into the interlayers ($Li_NTi_{16}S_{32}$, $1 \leq N \leq 15$) as the simulation cell, and sampled the Brillouin zone with a 2×2×4 $k$-point grid. The $ab$-axes lengths, $l_a$ and $l_b$, for the simulation cell are determined by $l_a = l_b = 0.01295N + 13.1954$, obtained by least-squares fitting to the $ab$-axes lengths obtained from structural relaxation at $N$=1,8,15, and 16. Temperature control was achieved using the Langevin thermostat with a friction coefficient of 5 THz. We set the MD time step to 2 fs, and the total simulation time was set to 400 ps (200,000 steps) for $N \leq 11$, whereas for $N \geq 12$ the simulation time was extended to 800 ps (400,000 steps).

## *Supplementary Materials*

In the RSE-MetaD simulations, the replica state exchanges are attempted every 100 MD steps. To reduce the enormous computational costs in the infinite swapping (INS) method, we employed the partial INS (pINS) method [25] for $N \geq 7$, in which parallel MetaD simulations are divided into multiple blocks, and INS is performed within each block. To approximate sampling from the full permutation space, two partition schemes with non-overlapping boundaries are used alternately. The partition schemes used in the present study are listed in Table SI, in which commas denote block boundaries, while each natural number indicates the length of each block. Figure S1 shows the example of the notation for $N = 12$, in which replicas are divided into the first three, the middle six, and the last three blocks (partition scheme A), and the first and the last six blocks (partition scheme B).

TABLE SI. The partition schemes in dual chain method used in the present study.

| $N$ | Partition scheme A | Partition scheme B |
|---|---|---|
| 7 | 1,6 | 6,1 |
| 8 | 2,6 | 6,2 |
| 9 | 3,6 | 6,3 |
| 10 | 4,6 | 6,4 |
| 11 | 2,6,3 | 5,6 |
| 12 | 3,6,3 | 6,6 |
| 13 | 2,5,6 | 5,6,2 |
| 14 | 3,6,5 | 6,5,3 |
| 15 | 3,6,6 | 6,6,3 |

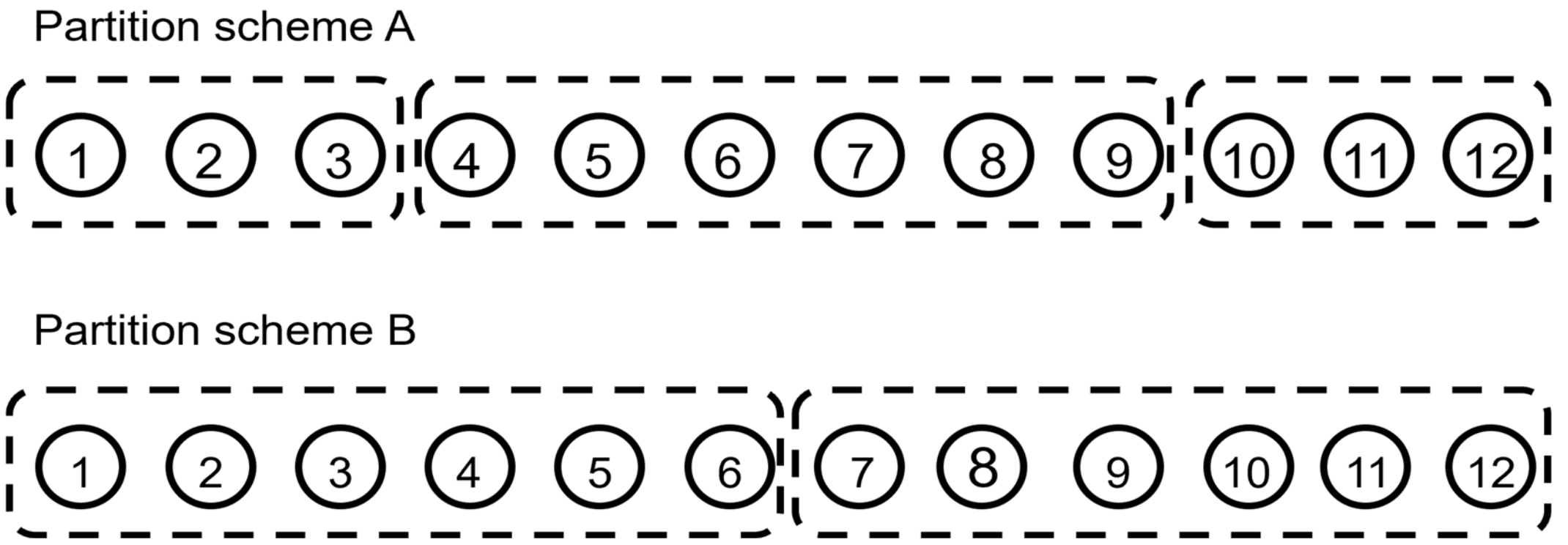

FIG. S1. Schematic illustration of the partition scheme for $N = 12$. The Numbered circles represent individual simulations, and the dashed lines indicate the blocks. These partition schemes correspond to those listed in Table SI, specifically (3,6,3) and (6,6) for schemes A and B, respectively.

## S2. The accuracy of machine learning force fields

Figure S2 shows the parity plots for (a), (c) the total energies and (b), (d) the atomic forces predicted by the machine learning force field (MLFF) during the RSE-MetaD and PB-MetaD simulations. Panels (a) and (b) correspond to $Li_{N/16}TiS_2$ ($2 \leq N \leq 15$) at 300 K, while panels (c) and (d) correspond to $Li_{11/16}TiS_2$ at 150, 200, 600, and 1500 K. For each concentration and temperature, 100 structures were sampled from each of the RSE-MetaD and PB-MetaD simulations, with the DFT calculation results plotted on the horizontal axis and the MLFF predictions on the vertical axis. In panels (b) and (d), the forces acting on all atoms are plotted together. The root-mean-square errors (RMSEs) of the predictions were calculated across all samples obtained from the RSE-MetaD and PB-MetaD simulations and are displayed on the plots. As shown in Fig. S2, the data points are distributed near the diagonal line across all Li compositions and temperatures. At 300 K, the RMSEs for total energies and atomic forces are within the ranges of 0.23–0.35 meV/atom and 23–30 meV/Å, respectively, indicating the high accuracy of the MLFF. Although the RMSEs tend to increase at higher temperatures, they remain within 2.0 meV/atom and 95 meV/Å, respectively, even at 1500 K.

# *Supplementary Materials*

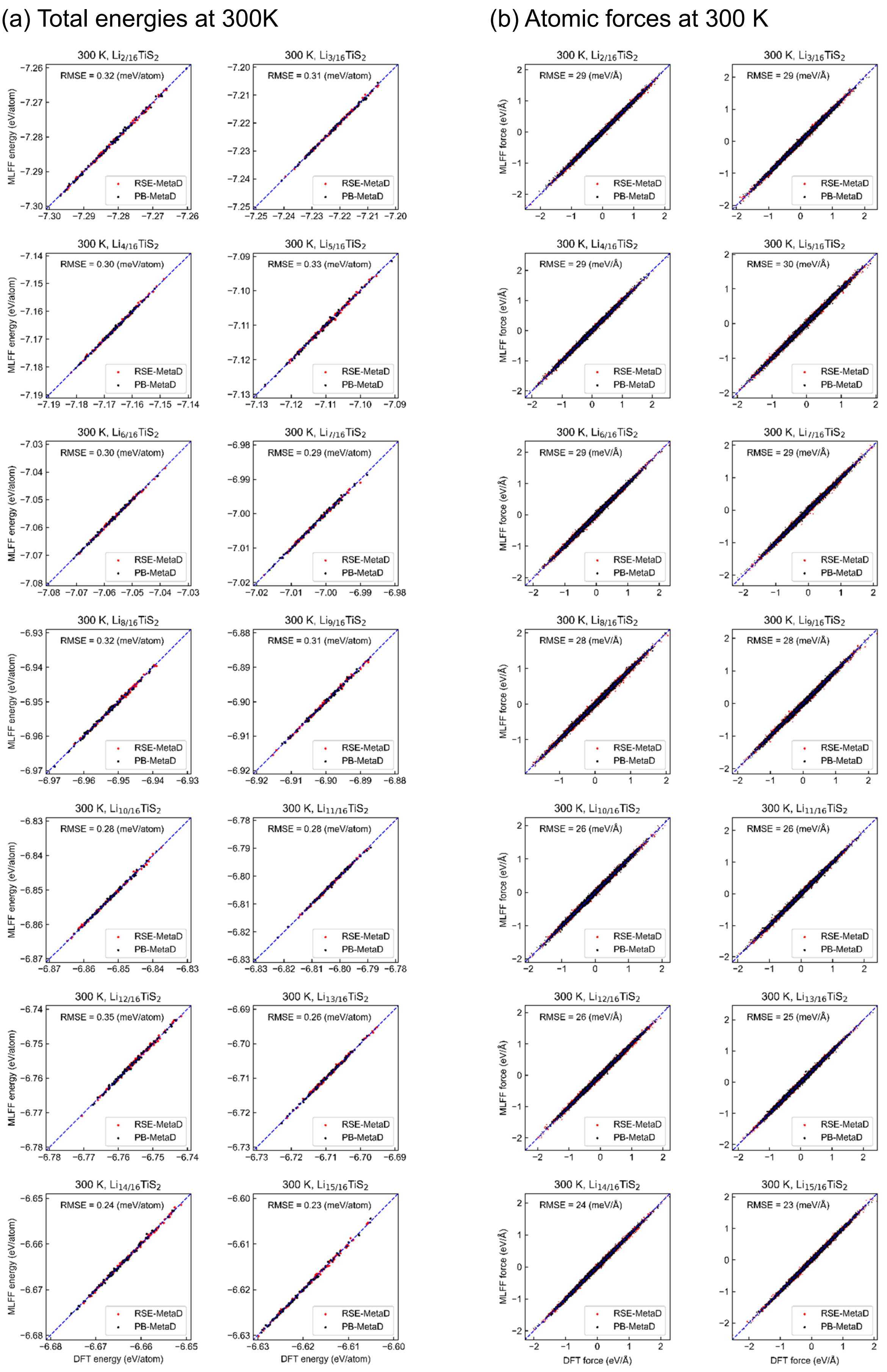

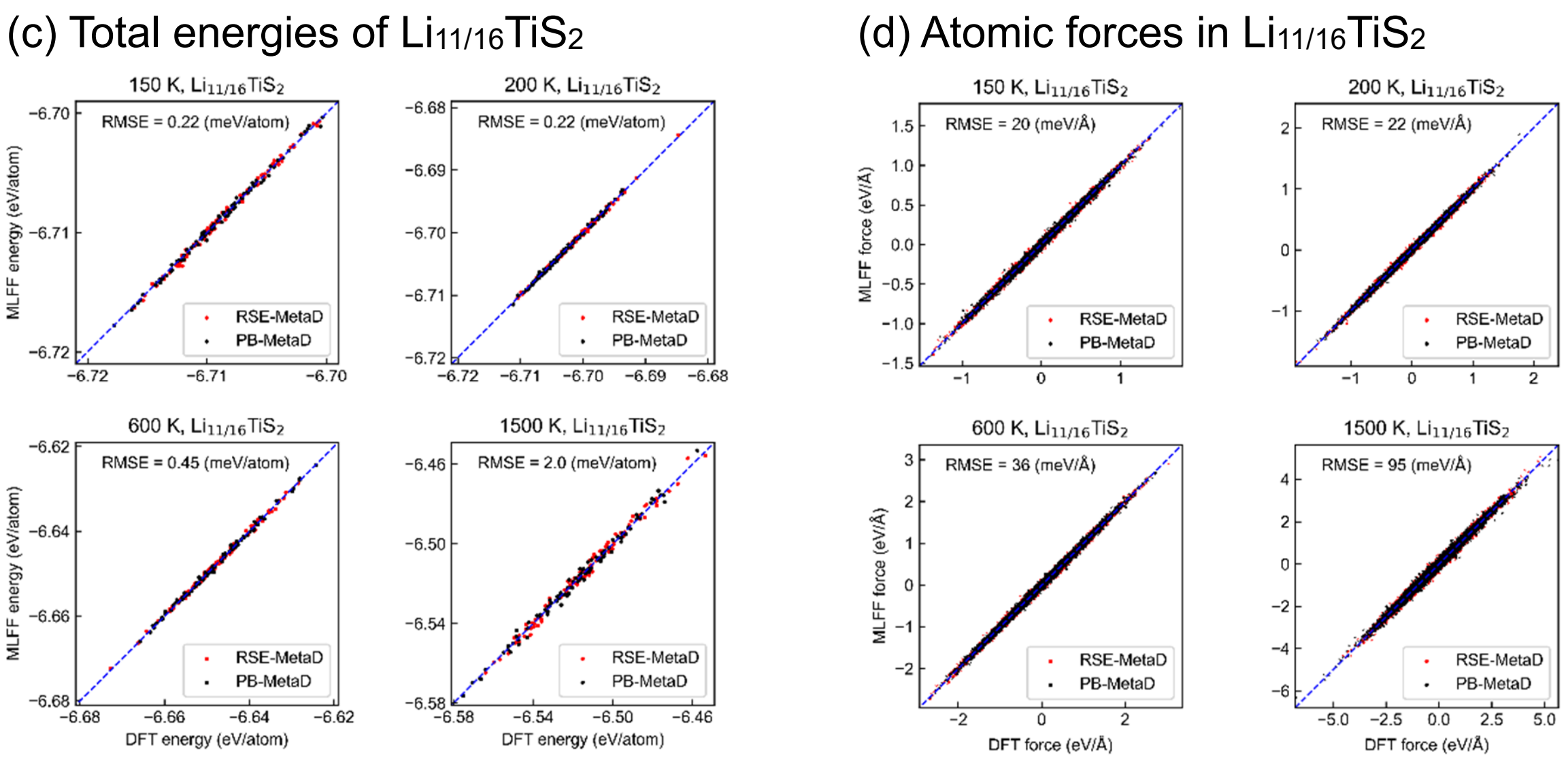

FIG. S2. Parity plots of the MLFFs employed in the RSE-MetaD and the PB-MetaD simulations. (a) Total energies and (b) atomic forces for $Li_{N/16}TiS_2$ ($N = 2, 3, \ldots, 15$) at 300 K. (c) Total energies and (d) atomic forces in $Li_{11/16}TiS_2$ at 150, 200, 600, and 1500 K.

**S3. Free energy surfaces over the full range of Li concentrations**

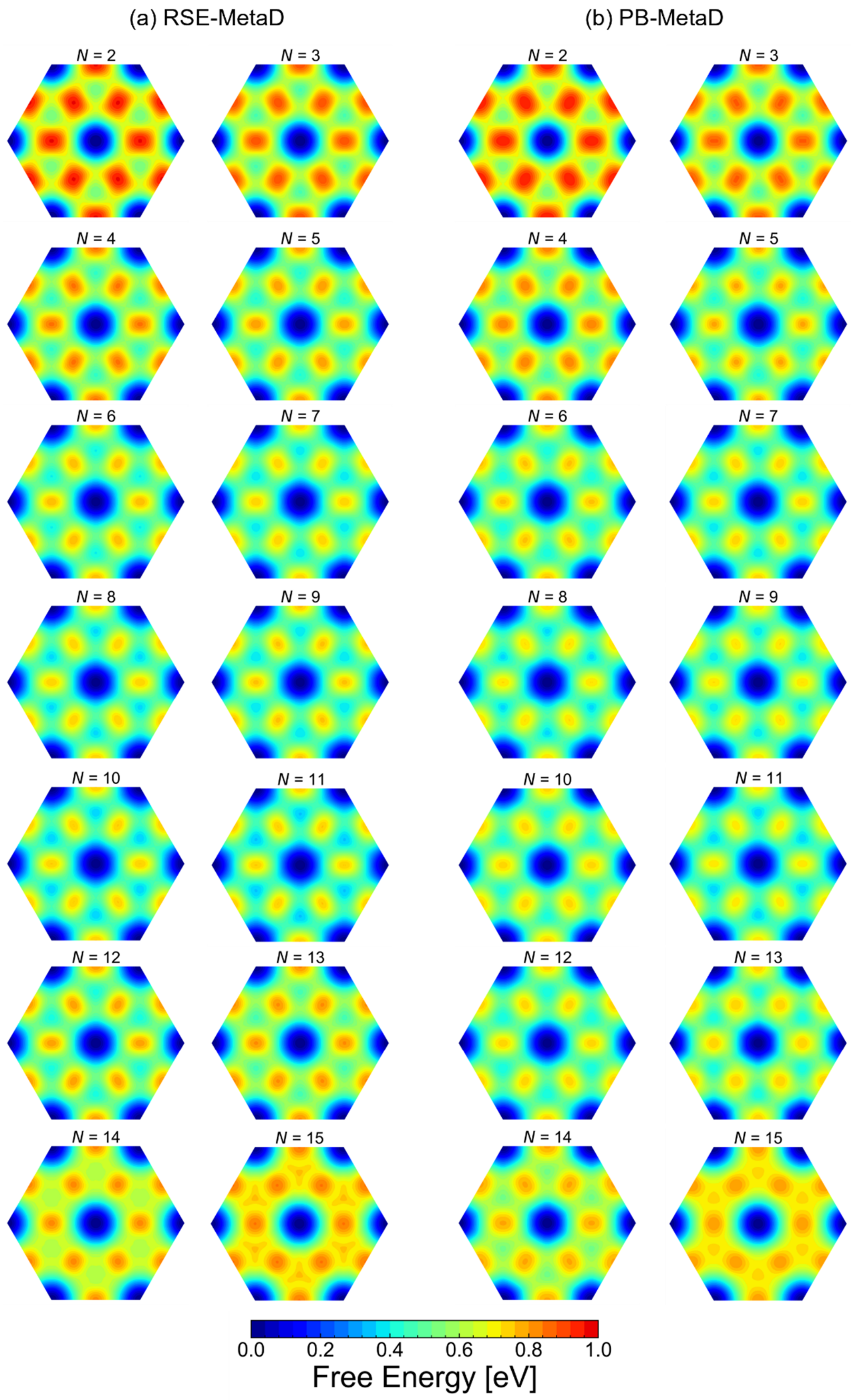

FIG. S3. Li-FESs in $Li_{N/16}TiS_2$ obtained from (a) RSE-MetaD and (b) PB-MetaD simulations. The color bar represents the free energy referenced to the most stable octahedral-site position.

**S4. Li-concentration dependence of the *c*-axis length**

Figure S4 shows the *c*-axis lattice parameter of $Li_xTiS_2$ as a function of Li concentration $x$. The grey down-triangle, red squares, black circles, blue diamonds, and green up-triangles represent the *c*-axis lengths averaged over multi-hill-MetaD, RSE-MetaD and PB-MetaD simulations at 300 K, those obtained from structural relaxation, and the experimental values from room-temperature XRD measurements [34], respectively. The *c*-axis lattice parameter remains nearly constant at high Li concentrations but decreases markedly for $x \leq 0.4$. This trend indicates that the interlayer spacing in $TiS_2$ correspondingly decreases with decreasing $x$ at the low Li concentration regime. The resulting contraction of interlayer spacing leads to an increase in the Li migration barrier due to enhanced electrostatic repulsion, resulting in an increase in the free energy barriers.

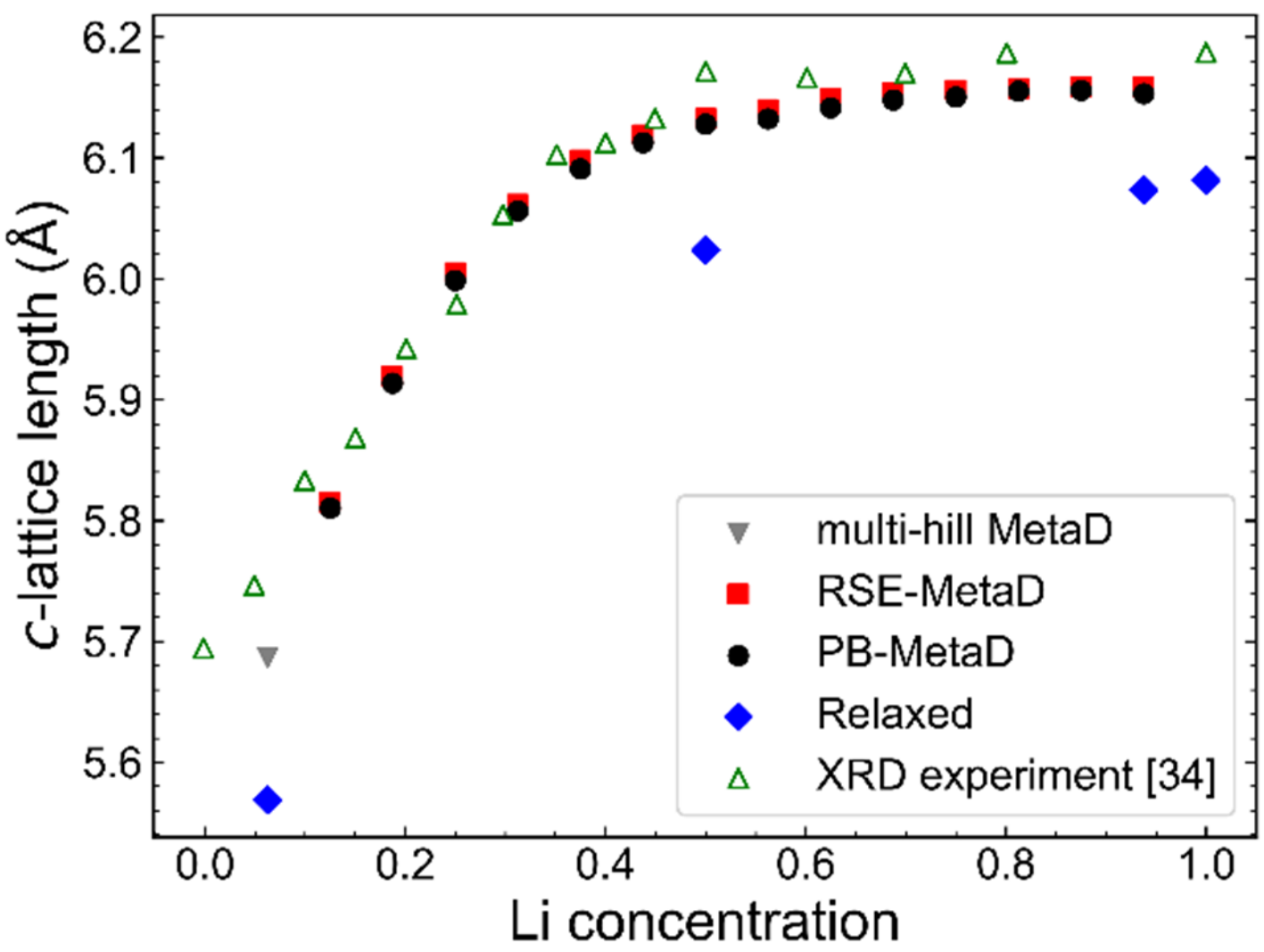

FIG. S4. *c*-axis lengths as a function of Li concentration. The grey down-triangle, red squares, black circles, blue diamonds, and green up-triangles represent the *c*-axis lengths averaged over multi-hill MetaD, RSE-MetaD and PB-MetaD, the optimal *c*-axis lengths by structural optimizations, and the experimental values from room-temperature XRD measurements [34], respectively.

## S5. Dynamical recrossing effects

To estimate the dynamical recrossing effects in $Li_xTiS_2$, we counted two types of migration processes in the standard MD simulation at 1500 K. The first is the total number of Li migration processes, $n_{tot}$, in which a Li atom jumps from an octahedral site to a tetrahedral site and subsequently to one of three adjacent octahedral sites. The second is the number of Li jumps returning to the original octahedral site from the tetrahedral site, $n_{back}$. We counted $n_{tot}$ and $n_{back}$ for divacancy and single-vacancy jumps separately (schematically illustrated in Fig. S5) to account for the different number of the final states; three and two for the divacancy and single-vacancy jumps, respectively. The line separating the octahedral and tetrahedral sites was defined as the straight line perpendicular to the minimum free energy path passing through the saddle point on the FES obtained by the PB-MetaD simulation at 1500 K. Note that the defined line separating the octahedral and tetrahedral sites are nearly identical on the FESs obtained by the RSE-MetaD and the PB-MetaD, which was located 1.34 Å from the center of the octahedral site.

In the MD simulation at 1500 K for 2 ns, $n_{tot}$ and $n_{back}$ are 7,798 and 2,151, for divacancy jumps, and 7,317 and 3,142, for single-vacancy jumps, respectively. Therefore, the transmission coefficients for divacancy and single-vacancy jumps, $\kappa_d$ and $\kappa_s$, were estimated to be 1.09 and 1.14, respectively, using the following equations:

$$\kappa_d = \frac{3}{2}\left(1 - r_{back,d}\right),$$

$$\kappa_s = 2\left(1 - r_{back,s}\right).$$

Here, $r_{back,d}$ and $r_{back,s}$ denote the ratio of returning to the original site, $n_{back}/n_{tot}$, for divacancy and single-vacancy jumps, respectively. The estimated $\kappa_d$ and $\kappa_s$ close to unity mean that the dynamical recrossing effect is negligible in this model system.

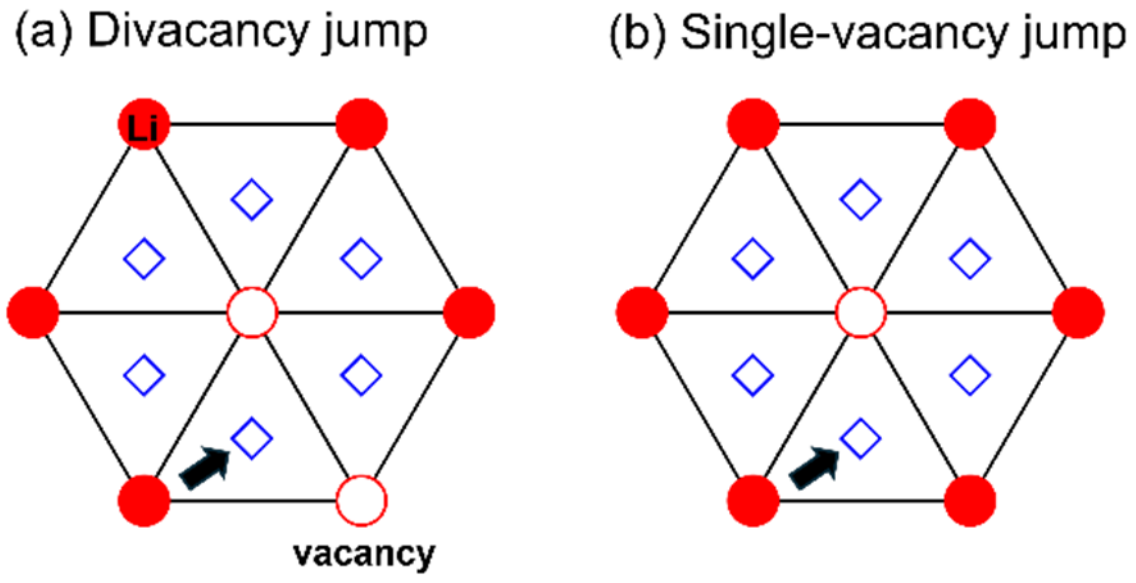

FIG. S5. Schematic illustration of (a) a divacancy jump and (b) a single-vacancy jump. Red circles and blue squares represent octahedral and tetrahedral sites, in which solid and open symbols correspond to occupied and vacant sites, respectively. Black arrows indicate the direction of Li migration from an octahedral site to a tetrahedral site. In a divacancy jump, a Li atom migrates into a tetrahedral site adjacent to both vacant octahedral sites, whereas in a single-vacancy jump, it migrates into a tetrahedral site adjacent to only a single vacancy.

**S6. Effect of *c*-axis length on the Li migration barrier**

As shown in Fig. S4 in Sec. S4, the mean *c*-axis lattice parameters obtained from both the RSE-MetaD and the PB-MetaD simulations are in good agreement with the experimental XRD values, whereas those obtained from structural optimizations are approximately 1–2% smaller. To evaluate the effect of this difference on Li migration barrier, we performed NEB calculations with the *c*-axis length constrained to different values.

Figure S6 shows the potential energy profiles along the Li migration pathway from an octahedral site to a neighboring tetrahedral site in (a) $Li_{1/16}TiS_2$ and (b) $Li_{15/16}TiS_2$. The *c*-lattice parameters were constrained to either the relaxed values or those accounting for the thermal expansion averaged over MetaD simulations at 300 K, corresponding to the multi-hill MetaD and PB-MetaD simulations for (a) and (b), respectively. The expanded *c*-axis constants reduce the Li migration barrier by 0.053 and 0.044 eV, corresponding to increases in the diffusion coefficient by factors of 7.8 and 5.5 at 300 K for $Li_{1/16}TiS_2$ and $Li_{15/16}TiS_2$, respectively.

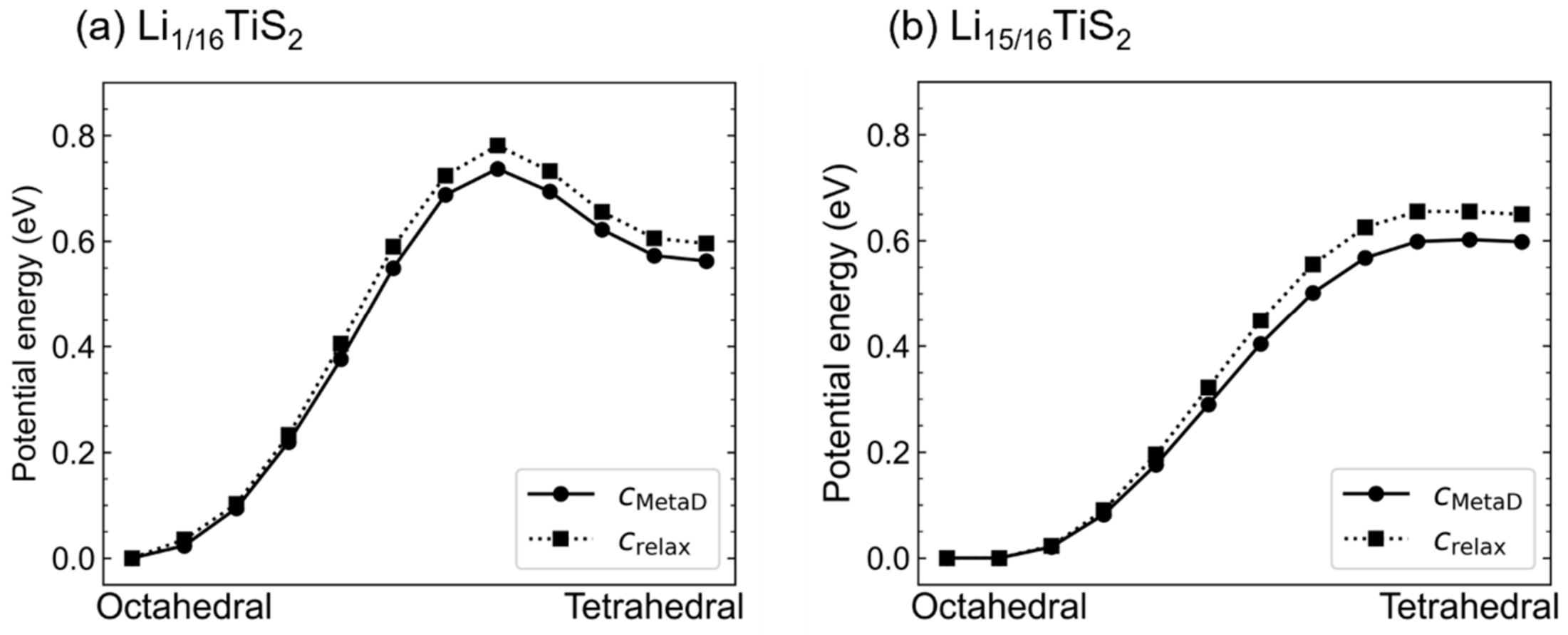

FIG. S6. Effects of *c*-axis lengths on Li migration barriers in (a) $Li_{1/16}TiS_2$ and (b) $Li_{15/16}TiS_2$. The circles and squares represent the results of NEB calculations with the *c*-axis constrained to the values averaged over the MetaD simulations $c_{\mathrm{MetaD}}$, and to the relaxed values $c_{\mathrm{relax}}$ from structural optimizations, respectively. $c_{\mathrm{MetaD}}$ and $c_{\mathrm{relax}}$ are 5.57 Å and 5.69 Å for $Li_{1/16}TiS_2$, and 6.08 Å and 6.15 Å for $Li_{15/16}TiS_2$, respectively. The NEB calculations were performed using 10 intermediate images and were converged until the residual force on each atom was below 0.01 eV/Å. The MLFFs constructed via the multi-hill MetaD or PB-MetaD simulations were employed for these calculations.

**S7. Dynamical correlation factors**

In the diffusion coefficients derived from single-particle FES using RSE-MetaD and PB-MetaD, the auto-correlation factor $f$ and the collective-correlation factor $f_c$ between atomic jumps are neglected. Therefore, we here examine whether rough estimation of these dynamical correlation factors is possible using the atomic trajectories in MetaD simulations, although they should ideally be estimated from the MD trajectories without an applied bias potential.

We enumerated all Li jumps between adjacent octahedral sites in the PB-MetaD simulations at 300 K for each Li concentration ($9/16 \leq x \leq 14/16$) in $Li_xTiS_2$. $f$ and $f_c$ can be estimated using the following definitional equations:

$$f = 1 + \frac{1}{N_{\mathrm{jump}}} \sum_{n=1}^{N_{\mathrm{Li}}} \sum_{i \neq j} \cos\theta_{ni,nj},$$

$$f_c = 1 + \frac{1}{N_{\mathrm{jump}}} \sum_{(n,i) \neq (m,j)} \cos\theta_{ni,mj}.$$

Here, $\theta_{ni,mj}$ represents the angle between the $i$-th jump vector of the $n$-th Li atom and the $j$-th jump vector of the $m$-th Li atom. $N_{\mathrm{Li}}$ and $N_{\mathrm{jump}}$ represent the number of Li atoms in the simulation cell and the total number of Li jumps in the single simulation, respectively. The Haven ratio $H_R$ can be obtained from $f$ and $f_c$ using the equation, $H_R = \frac{f}{f_c}$.

Figure S7 shows the estimated $f$, $f_c$, and $H_R$ in $Li_xTiS_2$ ($9/16 \leq x \leq 14/16$), in which $f$ and $f_c$ are in the ranges of 0.41–0.61 and 0.88–1.65, respectively. These are of a similar magnitude to the values for the random walk of a single vacancy on the two-dimensional hexagonal lattice, i.e., $f = 0.560061$ [39] and $f_c = 1$. $H_R$ ranges from 0.27 to 0.57, which deviates significantly from unity. The $H_R$ can also be estimated from the reported tracer diffusion coefficients $D^*$ and the jump diffusion coefficients $D_J$ in the literature based on the KMC simulations [12], i.e., $H_R = \frac{D^*}{D_J}$, which are shown by the open circles in the figure. Our estimated $H_R$ from PB-MetaD are in reasonable agreement with those estimated from the KMC in the literature, suggesting that these correlation effects can be roughly estimated through post-processing based on the information already obtained by PB-MetaD simulations.

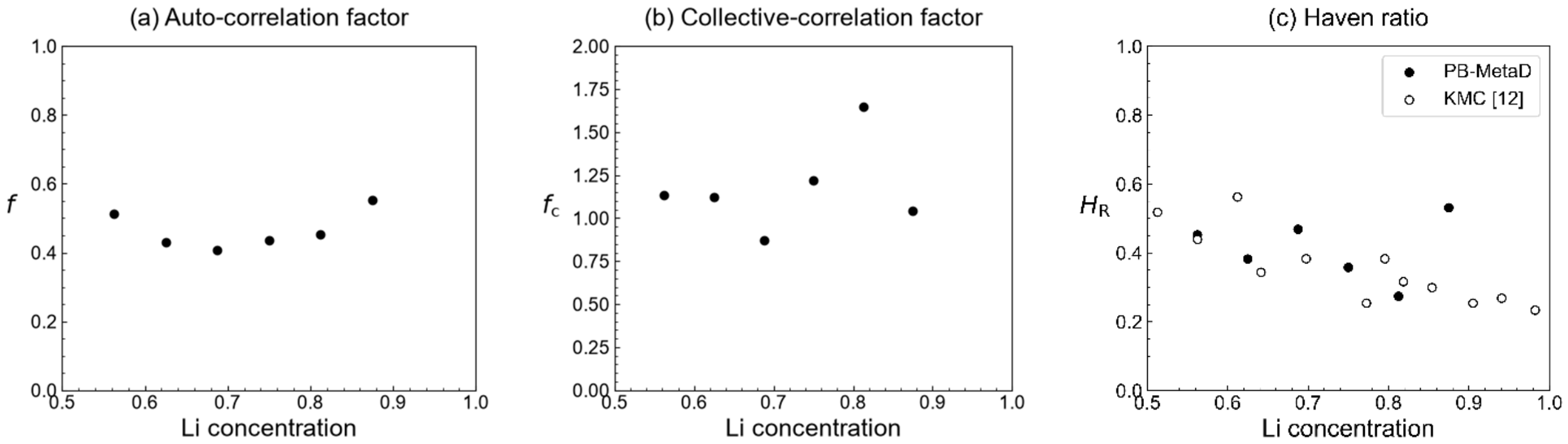

FIG. S7. (a) Auto-correlation factors, (b) collective-correlation factors, and (c) Haven ratios as a function of Li concentration estimated from the Li trajectories in the PB-MetaD simulations. In figure (c), $H_\mathrm{R}$ obtained from $D^*/D_\mathrm{J}$ in the previous KMC study [12] is also shown by open markers.

**S8. Convergence profiles of Li diffusion coefficients**

Figure S8 shows the convergence behavior of Li diffusion coefficients $D$ in $Li_{N/16}TiS_2$ ($N$=1,2,…,15) evaluated by RSE-MetaD (the red line) and PB-MetaD (the black line) methods. The $x$-axis denotes MD time steps and $y$-axis denotes the corresponding $D$ values obtained from the Gaussian hills accumulated up to each MD time step.

At the beginning of the simulations, few Gaussian hills are deposited, resulting in large $D$. As the Gaussian hills fill the free energy basin corresponding to the initial state, $D$ decreases dramatically and converges asymptotically. In $x \leq 11/16$, $D$ exhibits converged behavior in both RSE-MetaD and PB-MetaD. However, in $12/16 \leq x \leq 14/16$, $D$ remains fluctuated largely even in the final stage of the simulations, whereas at $x = 15/16$, the fluctuations become small again. The large fluctuation in $12/16 \leq x \leq 14/16$ can be attributed to the abrupt change of Li migration barrier between configurations with and without *divacancies*. Indeed, a previous DFT study [12] reported that the Li migration barrier via a divacancy is reduced by 0.25 eV compared with that via a single vacancy. When a divacancy is present, bias potentials are preferentially deposited around transition state due to the low Li migration barrier, resulting in low free energy barriers. In contrast, when a divacancy is absent, a relatively large number of bias potentials accumulates around the octahedral sites, resulting in high free energy barriers. These variations in the free energy barrier give rise to the fluctuations in the diffusion coefficients.

The blue lines in $9/16 \leq x \leq 14/16$ represent the results obtained from averaged FESs over $N$ independent PB-MetaD simulations. The fluctuations of $D$ profiles are considerably suppressed, indicating that the parallelization of PB-MetaD simulations is a solution to obtain a steady $D$ profile.

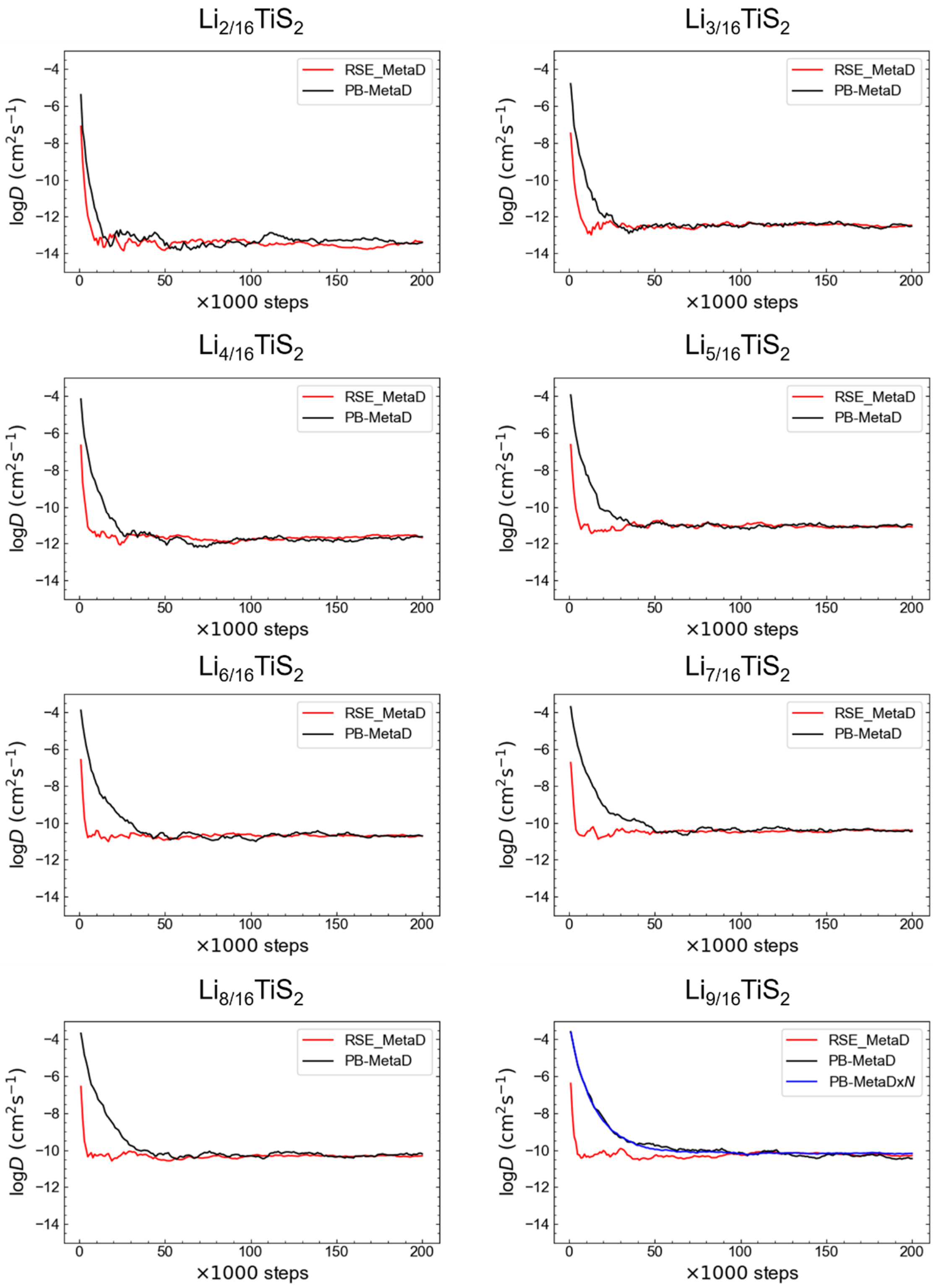
Li2/16TiS2
Li3/16TiS2
Li4/16TiS2
Li5/16TiS2
Li6/16TiS2
Li7/16TiS2
Li8/16TiS2
Li9/16TiS2
RSE_MetaD
PB-MetaD
PB-MetaDxN
logD (cm2s−1)
×1000 steps
−4
−6
−8
−10
−12
−14
0
50
100
150
200

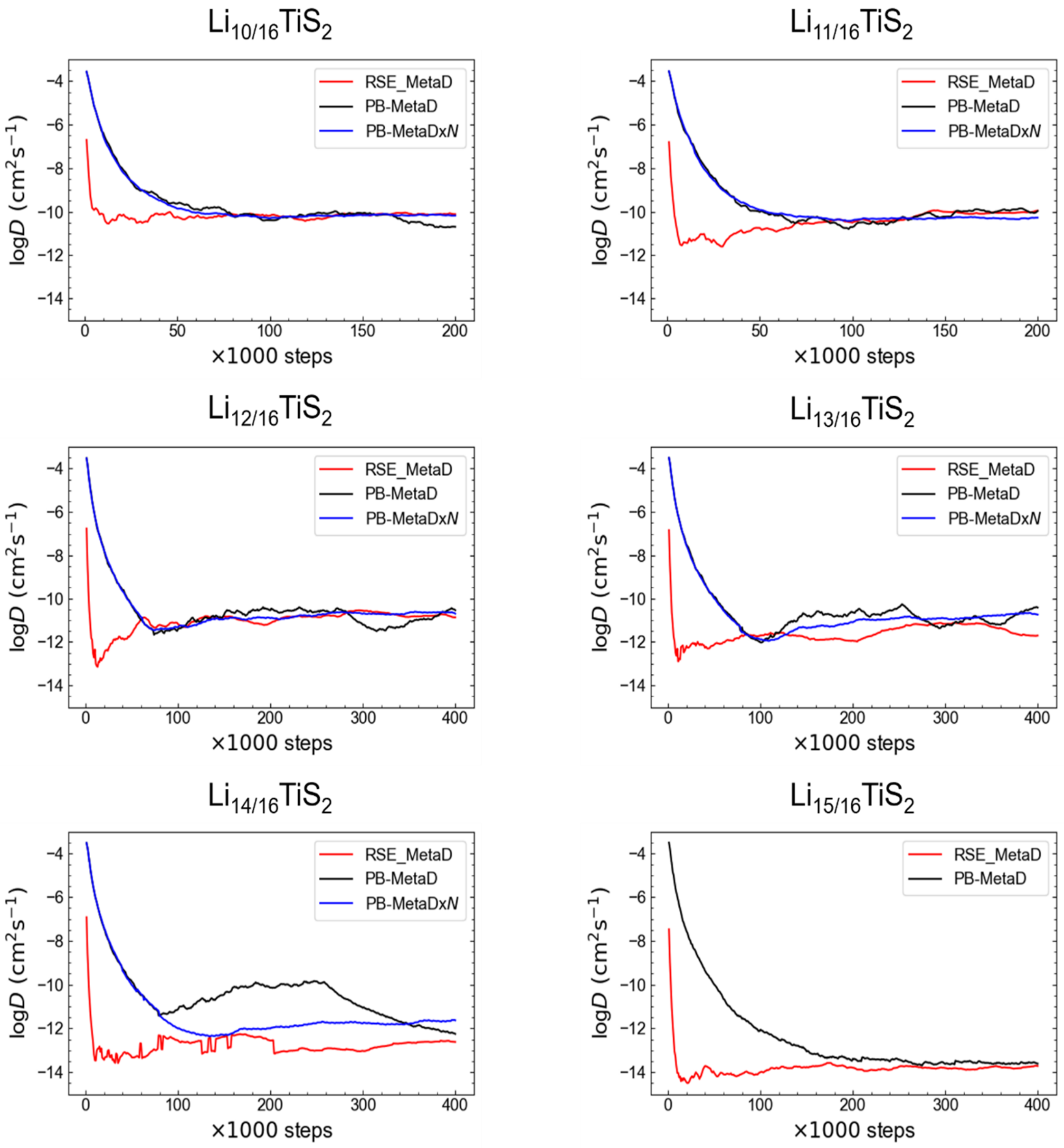

FIG. S8. Li diffusion coefficient profiles with respect to the MD time step. Red and black lines represent the profiles obtained from RSE-MetaD and PB-MetaD simulations, respectively. Blue lines for $9/16 \leq x \leq 14/16$ are obtained from averaged FES over $N$ independent PB-MetaD simulations. Diffusion coefficients are evaluated from the Gaussian hills accumulated up to each MD time step.

**S9. Discrepancy between the RSE-MetaD and PB-MetaD for $Li_{14/16}TiS_2$**

The underestimation of Li diffusion coefficients in $Li_{14/16}TiS_2$ by RSE-MetaD compared to PB-MetaD, is primarily attributed to the slow convergence of FES. This slow convergence is caused by persistent heterogeneity of local environments experienced by individual biased Li atoms, most notably regarding their adjacency to both vacancies forming a divacancy (hereafter referred to as "adjacency to the divacancy"). In principle, the biased Li atoms adjacent to the divacancy should be redistributed via replica state exchanges. However, such exchanges rarely occur due to the low selection probability, leaving the same biased Li atoms persistently adjacent to the divacancy.

To discuss the origin of the low selection probability, we consider a replica state exchange from a state $\boldsymbol{\sigma} = (\sigma_1, \sigma_2, \ldots, \sigma_N)$ to $\boldsymbol{\sigma}' = (\sigma'_1, \sigma'_2, \ldots, \sigma'_N)$ ($\boldsymbol{\sigma}' \neq \boldsymbol{\sigma}$), the statistical weight of which is given by Eq. (5) in the main text. In general, the statistical weight $W_{\boldsymbol{\sigma}'}^{\mathrm{RSE}}$ tends to be lower than $W_{\boldsymbol{\sigma}}^{\mathrm{RSE}}$, meaning that the identity operation has a high selection probability. This is because the coordinate of the $i$-th carrier in $\sigma_i$ tend to deviate from the stable site due to the bias potential, whereas the $i$-th carrier in $\sigma'_i$ likely to be in the vicinity of the stable site, resulting in

$$V_i^{\mathrm{RSE}}\left(\boldsymbol{X}_{\sigma'_i}^{(i)}\right) > V_i^{\mathrm{RSE}}\left(\boldsymbol{X}_{\sigma_i}^{(i)}\right).$$

If the $i$-th Li atom in $\sigma_i$ is adjacent to the divacancy, the deviation from the stable site is enlarged due to the reduced potential barrier from the octahedral site to the tetrahedral site, leading to a further reduced $W_{\boldsymbol{\sigma}'}^{\mathrm{RSE}}$. Consequently, this suppressive effect explains why replica state exchanges involving divacancy-adjacent Li atoms are rarely selected.

Figure S9 shows the occurrence of the persistent adjacency during the RSE-MetaD simulation, where the upper red bands represent the period during which the biased Li atom is adjacent to the divacancy. The lower black lines denote the observations of replica state exchanges. Figure S9 suggests that replica state exchanges occur frequently while the biased Li atom is not adjacent to the divacancy, whereas the presence of such adjacency suppresses the exchanges.

To quantify the effects of the exchange suppression, we introduce divacancy adjacency rate for each biased Li atom $r_{\mathrm{div}}$ defined as

$$r_{\mathrm{div}} = \frac{\Delta t_{\mathrm{div}}}{t_{\mathrm{sim}}}$$

where $t_{\mathrm{sim}}$ and $\Delta t_{\mathrm{div}}$ is the total simulation time and the total time interval during which the biased Li atom is adjacent to the divacancy, respectively. Note that a single Li atom is biased in a simulation in RSE-MetaD, whereas all Li atoms are biased in a single simulation in PB-MetaD.

Figure S10 shows the time evolution of $r_{\mathrm{div}}$ alongside the corresponding FESs $F_i^{\mathrm{SP}}(\boldsymbol{X})$ for (a) RSE-MetaD and (b) PB-MetaD. $F_i^{\mathrm{SP}}(\boldsymbol{X})$ were reconstructed from the Gaussian hills accumulated on each biased Li atom at the end of the simulations via

$$F_i^{\mathrm{SP}}(\boldsymbol{X}) = -\left(1 + \frac{T}{\Delta T}\right) V_i(\boldsymbol{X}),$$

where the symbols are defined in Eqs. (6) and (12) in the main text. In the RSE-MetaD, the time evolution of $r_{\mathrm{div}}$ clearly differs among individual biased Li atoms; some remain adjacent to the divacancy for more than 30% of the total simulation time, whereas others do not experience such adjacency at all during the 800 ps run. Consequently, the Li atoms with high $r_{\mathrm{div}}$ yield FESs with deep local minima at the tetrahedral metastable sites and low free energy barriers from the octahedral sites to the tetrahedral sites, whereas those with low $r_{\mathrm{div}}$ result in shallower minima and higher free energy barriers. On the other hand, in the PB-MetaD simulation, $r_{\mathrm{div}}$ exhibits similar behaviors across all individual Li atoms, resulting in the homogeneous $F_i^{\mathrm{SP}}(\boldsymbol{X})$ profile. This is because individual biased Li atoms experience adjacency to the divacancy through the divacancy diffusion, since all Li atoms are biased simultaneously in the PB-MetaD simulation. The difference in the homogeneity of $F_i^{\mathrm{SP}}(\boldsymbol{X})$ between the RSE-MetaD and the PB-MetaD simulations affects the resulting $F^{\mathrm{SP}}(\boldsymbol{X})$. Specifically, in RSE-MetaD, the $F_i^{\mathrm{SP}}(\boldsymbol{X})$ with higher free energy barriers contribute to increased free energy barriers, thereby reducing the diffusivity of Li atoms compared to the PB-MetaD simulation.

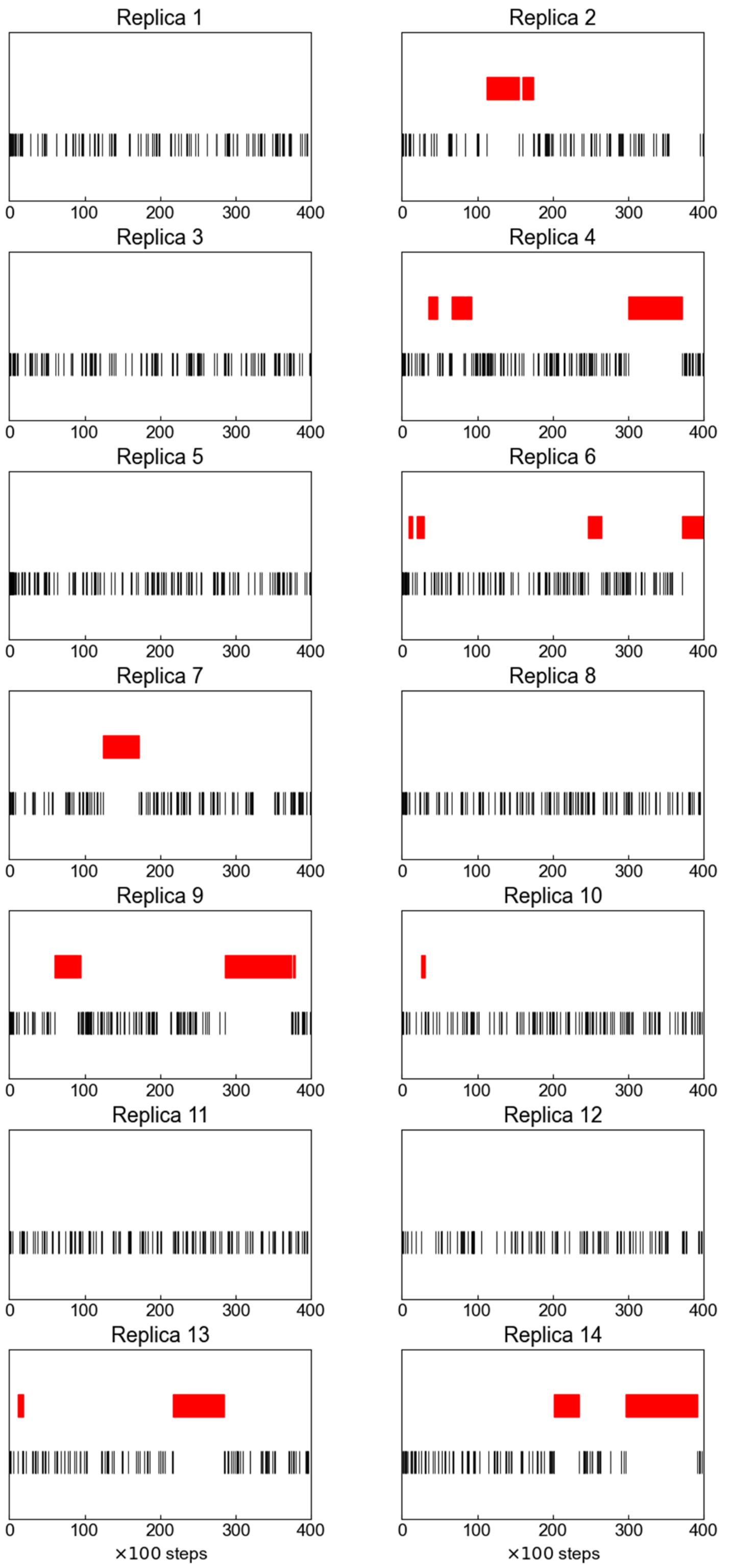

FIG. S9. Divacancy adjacency duration and the occurrence of replica state exchanges. The upper red bands represent the periods during which the biased Li atom was adjacent to the divacancy, while the lower black lines indicate the MD steps in which the replica state exchanges occurred.

(a) RSE-MetaD

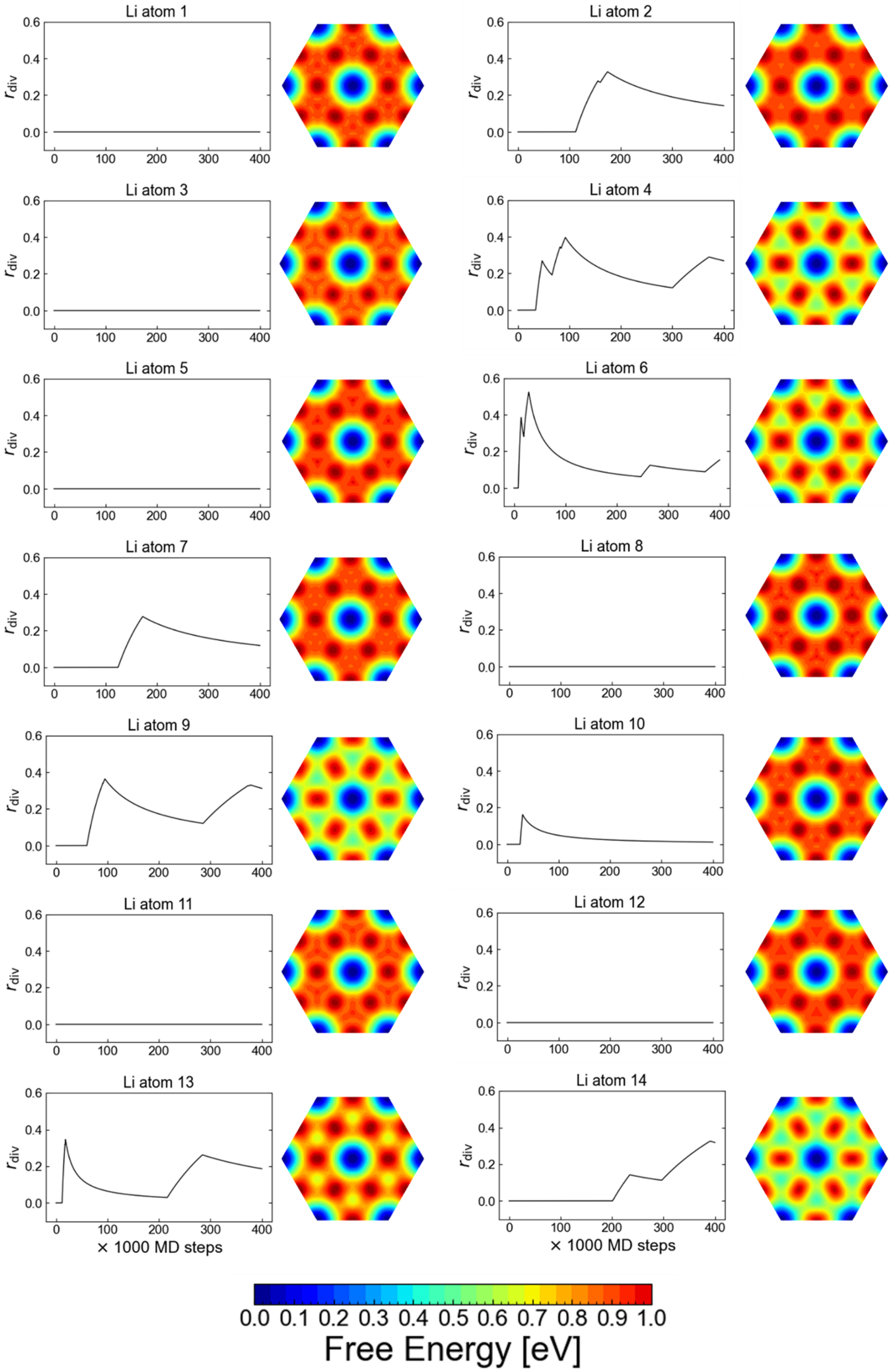
Li atom 1
Li atom 2
Li atom 3
Li atom 4
Li atom 5
Li atom 6
Li atom 7
Li atom 8
Li atom 9
Li atom 10
Li atom 11
Li atom 12
Li atom 13
Li atom 14
$r_{div}$
0.0
0.2
0.4
0.6
0
100
200
300
400
× 1000 MD steps
0.0 0.1 0.2 0.3 0.4 0.5 0.6 0.7 0.8 0.9 1.0
Free Energy [eV]

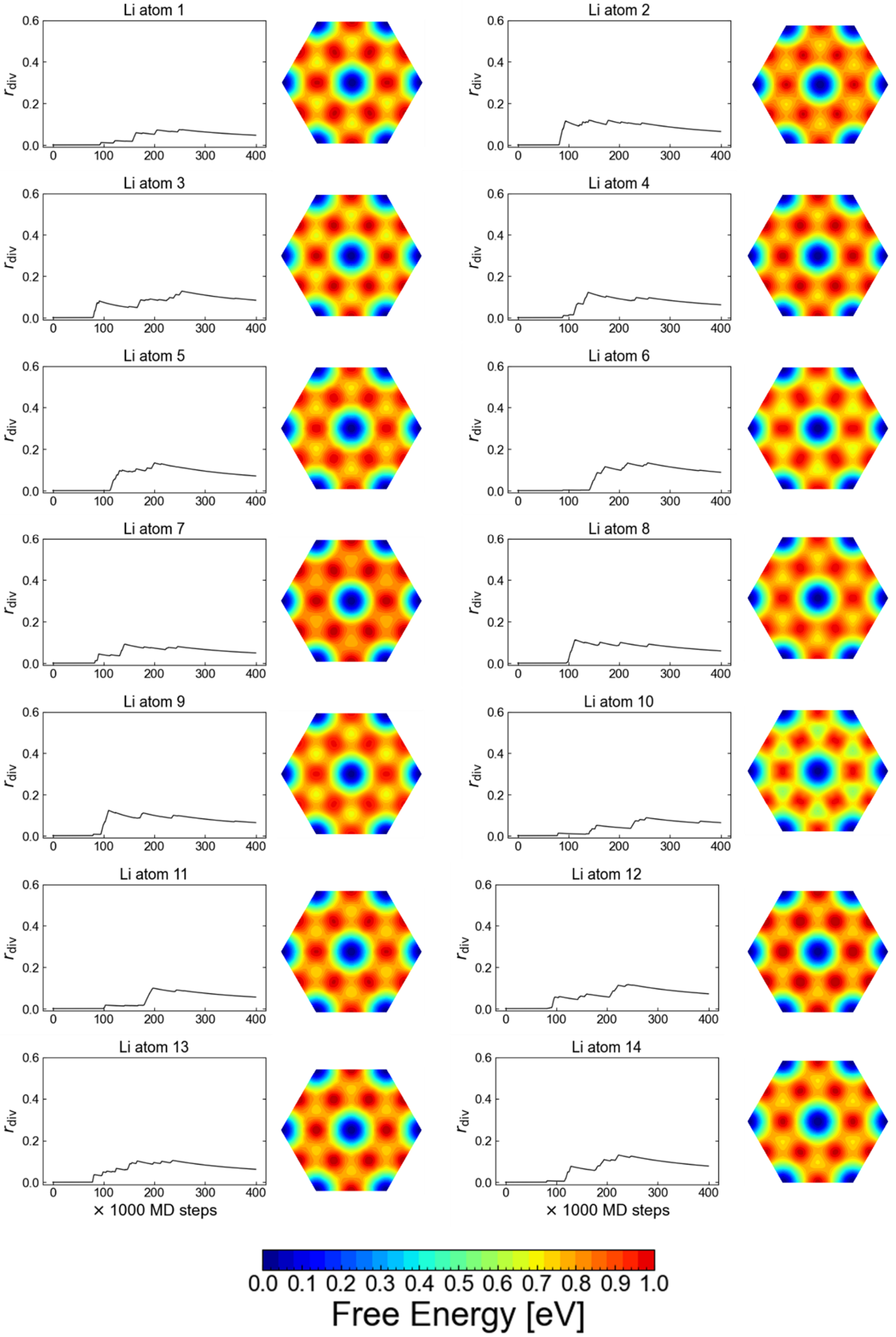

FIG. S10. Divacancy adjacency rates $r_{\mathrm{div}}$ and the resulting FESs for individual biased Li atoms in (a) the RSE-MetaD and (b) the PB-MetaD simulations.

**S10. Effect of temperature on the convergence profile of Li jump frequencies**

Figure S11 shows the convergence profiles of the Li jump frequency $\Gamma$ in $Li_{11/16}TiS_2$ from the octahedral sites to the tetrahedral sites as evaluated by RSE-MetaD (red lines) and PB-MetaD (black lines) at 150, 200, 300, 600, and 1500 K. The *x*-axis denotes MD time steps and *y*-axis denotes the corresponding $\Gamma$ values obtained from the bias potential accumulated up to each MD time step. Figure S11 indicates that $\Gamma$ converges well by 200,000 MD time steps at all temperatures examined, suggesting that the computational costs for RSE-MetaD and PB-MetaD are not strongly affected by temperature, unlike in standard MD simulations.

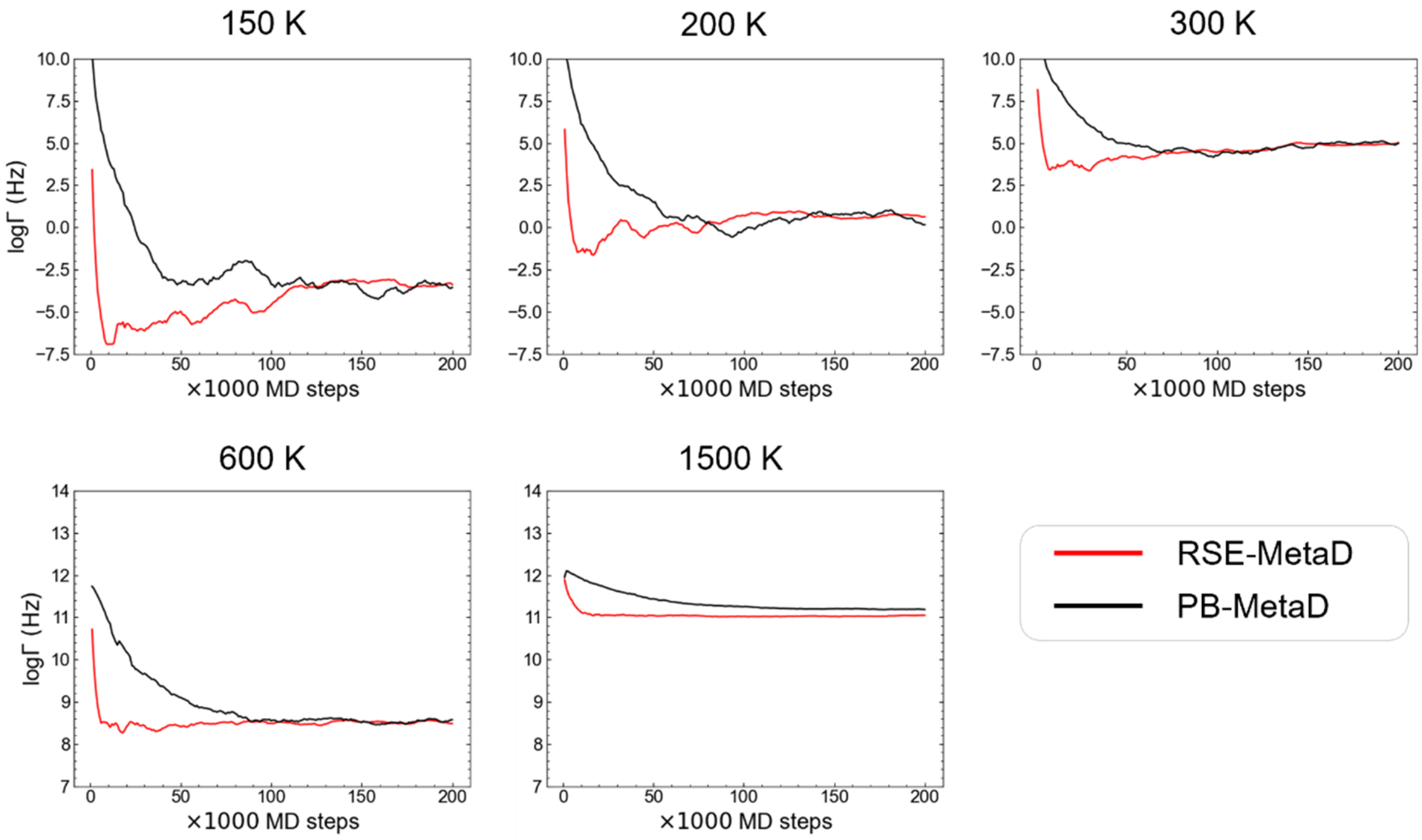

FIG. S11. Li jump frequency profiles from octahedral to tetrahedral sites in $Li_{11/16}TiS_2$ at various temperatures. The red and black lines represent the profiles obtained from the RSE-MetaD and the PB-MetaD simulations, respectively. The *x*- and *y*-axes denote the MD time steps and the jump frequency evaluated from the Gaussian hills accumulated up to each MD time step, respectively.